\documentclass[twocolumn]{aastex631}
\usepackage{amsmath}

\usepackage{soul}
\usepackage{subfigure}
\usepackage{graphicx}

\begin{document}

\title{
Decoding AGN Feedback with X-arithmetic: From Morphology to Physical Mechanisms
}

\author[0000-0003-3537-3491]{Hannah McCall}
\affiliation{Department of Astronomy and Astrophysics, University of
Chicago, 5640 S Ellis Ave, Chicago, IL 60637, USA}

\author{Irina Zhuravleva}
\affiliation{Department of Astronomy and Astrophysics, University of
Chicago, 5640 S Ellis Ave, Chicago, IL 60637, USA}

\author[0000-0002-0322-884X]{Eugene Churazov}\affiliation{Max Planck Institute for Astrophysics, Karl-Schwarzschild-Str. 1, D-85741 Garching, Germany}

\author[0000-0001-5888-7052]{Congyao Zhang}\affiliation{Department of Astronomy and Astrophysics, University of
Chicago, 5640 S Ellis Ave, Chicago, IL 60637, USA}\affiliation{Department of Theoretical Physics and Astrophysics, Masaryk University, Kotlářská 267/2, 61137 Brno, Czech Republic}

\author[0000-0002-9478-1682]{William Forman}\affiliation{SAO-CfA, 60 Garden St, Cambridge, MA 02138, USA}

\author[0000-0003-2206-4243]{Christine Jones}\affiliation{SAO-CfA, 60 Garden St, Cambridge, MA 02138, USA}

\author{Yuan Li}\affiliation{Department of Astronomy, University of Massachusetts Amherst, 710 N Pleasant St, Amherst, MA 01002, USA}

\begin{abstract}

Feedback from Active Galactic Nuclei (AGN) is a key process in the evolution of massive halos in the Universe. 
New observational information on feedback is crucial for improving the implementation of the physics in numerical models. 
In this work, we apply a novel image-manipulation technique, {termed} `X-arithmetic', to a sample of 15 galaxy clusters and groups deeply observed with Chandra. 
This technique decomposes perturbations in feedback-dominated regions into images excluding either (1) weak shocks and sound waves, (2) bubbles inflated by jets, or (3) cooling and slow gas motions (isobaric perturbations), enabling efficient spatial identification of these features without involving spectroscopic analysis. 
We confirm the nature of previously (spectroscopically-)identified features and newly establish the origin of other structures. 
We find that feedback produces multiple shocks in groups and massive galaxies, but only one to two shocks in clusters. 
Prominent isobaric structures are abundant around inner cavities in clusters, compared to almost no such structures in groups. 
These differences suggest that feedback effects are stronger in smaller-mass systems, possibly due to the shallower gravitational potential of groups or more violent feedback.
Follow-up spectroscopy, guided by the X-arithmetic results, suggests that earlier-identified “isothermal shocks” could be a mix of isobaric and adiabatic structures. 
We applied X-arithmetic to {galaxy cluster simulations}, demonstrating {its straightforward application and future potential for testing the feedback physics details in simulations}.
Our feasibility study shows that imaging data from future X-ray observatories like AXIS will be ideal for expanding X-arithmetic application to a larger sample of objects.

\end{abstract}


\section{Introduction} \label{sec:intro}

It is well established that AGN feedback is the key mechanism for quenching star formation in the most massive halos in the Universe, such as giant elliptical galaxies, galaxy groups, and clusters  \citep[see, e.g.,][]{Boehringer_1993, Churazov_2000, churazov_2002, Werner_2019, hlavacek_2022}. 
In particular, ``radio mode'' AGN feedback, which exhibits a low accretion rate and relativistic jets and/or radio lobes \citep[e.g.,][]{Birzan_2020}, is most relevant to present-day massive systems. 
During this process, AGN inflate bubbles of relativistic plasma \citep{Boehringer_1993, Churazov_2000, mcnamara_2000}. 
Their initial rapid inflation drives weak shocks and sound waves, which seem to play an important role in heating gas both early in an outburst and in lower-mass gaseous halos \citep[e.g.,][]{Forman_2007, Graham_2008,nulsen_2013, Randall_2015}. In many cluster-mass halos, after the initial eruption, the majority of the energy goes into the enthalpy of the inflated bubbles \citep[e.g.,][]{Hlavacek-Larrondo2015}.
During the subsequent rise of the bubbles, the bubble enthalpy is released to the ambient gas \citep{Churazov_2001,Begelman_2001}. 
Various physical processes are involved in the energy transfer, including generation of turbulence in the bubble wake, excitation of internal gravity waves, and uplift of the low entropy gas entrained by the rising bubbles \citep[e.g.,][]{Churazov_2001,Zhang_2018,Balbus_1990}. 
Some of these processes could trigger additional cooling of the gas, which could accrete onto the black hole, further powering the jets \citep[e.g.,][]{Pizzolato_2005, Hillel_2014}. 
In addition, cosmic ray proton streaming \citep[e.g.,][]{guo_2008, Pfrommer_2013} and mixing of the bubble gas with the intracluster medium (ICM) \citep[e.g.,][]{hillel_2020} have been proposed as possible energy-transferring mechanisms. 
Cosmic ray heating is only capable of balancing cooling at small radii, however, and requires the presence of thermal conduction at larger radii \citep{Jacob_2017}, while magnetic draping may interfere with the disruption of the bubbles that is required for efficient mixing \citep{Dursi_2008}. 
More recently, it was shown that enhanced scattering of cosmic rays, due to plasma mirror instabilities generated in the ICM, could confine sub-TeV cosmic rays within bubbles, supporting their observed integrity \citep{Ewart_2024}. 
In the context of isolated cluster atmospheres or cosmological simulations, AGN feedback has been studied numerically, reproducing observational results with various levels of success \citep[e.g.,][]{Li14, Li15, Reynolds_2015, Yang_2016, Zhang_2022, Ehlert_2023, Weinberger_2023}.

Due to the presence of hot gaseous atmospheres, many of these processes are imprinted on X-ray images, allowing the study of AGN feedback physics. 
For instance, if the duration of bubble inflation is short (rapid or violent inflation), most energy from the AGN will be transferred to strong shocks and shock-heated gas, leaving rather small cavities. 
In the opposite scenario of slow (long duration or gentle) bubble inflation, most energy is released in bubbles, creating comparatively large cavities, weak shocks, and weakly shocked gas \citep{Forman_2017}. There are also observable distinctions resulting from the power of the jet and the environment around the black hole. 
When jets are high-powered, they break through the ICM and propagate to large distances before interacting with the surrounding medium, creating backflows as seen in Fanaroff-Riley (FR) class II sources. 
Lower density surroundings allow the jet to reach larger distances. 
In the scenario where jets are less powerful, they become kink-unstable within the cores and stall, resulting in the inflation of large cavities filled with relativistic plasma as in FR I sources \citep[][]{Tchekovskoy_2016}.

To distinguish between different scenarios {for bubble inflation and jet propagation}, one relies on the identification of various structures present in the gas. 
X-ray cavities associated with bubbles filled with relativistic particles and emitting radio synchrotron emission have been reliably found in many sources \citep[e.g.,][]{Dunn_2006, Fabian_2012,Birzan_2004,Hlavacek-Larrondo_2012}. 
In addition, many structures have been identified and named based on their visual appearance \citep[e.g., hooks, plumes, bays, rings, arcs, channels, filaments, wings, to name a few, see., e.g.,][]{Blanton_2011, Sanders_2016, Forman_2007, Tremblay_2012, Randall_2010, Randall_2015}. With follow-up X-ray spectroscopic studies, some of these structures (mostly edges) were found to be shocks and cold fronts. 
These identifications often require much experimentation and meticulous analysis, including spectroscopic deprojection.

It is possible, however, to optimize the analysis and understand physical processes responsible for creating these surface brightness structures by simply manipulating X-ray images in different energy bands. 
The ``X-arithmetic" method, which takes advantage of this fact, was introduced in \cite{churazov_2016} and applied to just two clusters so far, Virgo and Perseus. 
In this paper, we apply X-arithmetic to a wider sample of 15 clusters, groups, and massive galaxies observed deeply ($> 200$\,ks) with Chandra, demonstrating the efficacy of the method in providing fast and accurate identification of the origin of bright features in the regions affected by AGN feedback\footnote{Although focused here on bright structures within cool cores, the X-arithmetic method can be applied to any structures within the ICM.}.
This increased sample also allows us to identify trends in types of features present, and, therefore, dominant processes driving AGN feedback in objects across a range of masses and radio properties. With X-arithmetic, we can readily make the leap from morphological classifications to physical ones.

The structure of this paper is as follows: in Section 2, we report our sample and explain its selection. Section 3 includes a description of the data processing and image preparation, while a full description of the X-arithmetic method comes in Section 4. In Section 5 we present our results and identified structures. Section 6 contains a follow-up spectroscopic investigation of new features and a discussion of the prospects for applying this method to cosmological simulations and data from future instruments. We summarize our findings in Section 7.

Throughout this paper, we assume the relative solar abundance of \cite{Anders_1989} and use \texttt{XSPEC} version 12.11.1 \citep{Arnaud_1996} and AtomDB version 3.0.9. We adopt the cosmological parameters $H_0 = 70\,\mathrm{km\ s}^{-1} \mathrm{Mpc}^{-1},\ \Omega_M=0.3,\ \mathrm{and\ } \Omega_\Lambda=0.7$.

\section{Sample Selection} \label{sec:sample}

\begin{table*}
    \centering
    \begin{tabular}{lcccccccccc}
       \hline
       Object & T & t$_{\mathrm{exp}}$ & \textit{z} & n$_{\rm H}$ & Soft & Hard & Model & Patching & Abund. & T$_0$\\
        & keV & ks &  & $\times 10^{20} \mathrm{cm}^{-2}$ & keV & keV &  & arcsec & Z$_{\odot}$ & keV\\
       \hline
       M84 & 0.6 & 852 & 0.004 & 2.90 & 0.5-1.0 & 1.0-6.5& SE & 20& 0.50 & 1.0\\
       NGC 5813 & 0.7 & 638 & 0.0064 & 4.26 & 0.5-1.0 & 1.0-7.5& DE & 50 & 0.40 & 0.7\\
       NGC 5044 & 1.0 & 398 & 0.009 & 4.90 & 0.5-1.3 & 1.3-7.0& DS & 90& 0.50 & 1.0\\
       M87/Virgo & 2.2 & 1238 & 0.004 & 2.50 & 0.5-3.5 & 3.5-7.5& DE & 100 & 0.75 & 1.0\\
       A2052 & 3.0 & 617 & 0.035 & 2.71 & 0.5-3.5 & 3.5-8.0& DE & 80 & 0.50 & 1.5\\
       Centaurus & 3.0 & 669 & 0.010 & 12.0 & 0.5-3.5 & 3.5-8.0& DE & 60& 0.50 & 1.2\\
       A133 & 3.0 & 197 & 0.057 &1.45 &  0.5-2.5 & 2.5-7.5& DE & 100& 1.0 & 2.5\\
       A3847 & 3.0 & 217 & 0.153 & 2.06 & 0.5-3.0 & 3.0-7.5& DE & 100& 0.30 & 1.0\\
       Hydra A & 3.5 & 190 & 0.054 & 4.00 & 0.5-3.0 & 3.0-7.5& SE & 90& 0.40 & 2.4\\
       A1795  & 3.5 & 3120 & 0.063 & 1.00 & 0.5-3.5 & 3.5-7.5& DE & 80& 0.70 & 3.5\\
       A2597 & 3.5 & 586 & 0.085 & 2.48 & 0.5-3.5 &3.5-7.5 & DE & 50& 0.50 & 2.75\\
       Phoenix & 4.0 & 546 & 0.597 & 1.43 & 0.5-3.5 & 3.5-8.0& DS & 0& 0.60 & 3.0\\
       Perseus & 4.0 & 1606 & 0.0175 & 14.6 & 0.5-3.5 & 3.5-7.5 & DE & 90 & 0.50 & 3.0 \\
       MS 0735.6+7421 & 5.0 & 526 & 0.216 & 3.10 & 0.5-3.5 & 3.5-7.5& DE & 100 & 0.40 & 3.5 \\
       Cygnus A & 5.0 & 795 & 0.056 & 19.0 & 0.5-3.5 & 3.5-7.5& DE & 30& 0.66 & 3.0\\
     \hline
    \end{tabular}
    \caption{The sample of objects: their names, representative temperatures (value that best describes gas temperature as a whole), cleaned exposure times available in the Chandra archive (ACIS-I/S, observed after 2001, small offsets from target center), redshift, the HI column density, soft and hard energy bands, underlying model type, patching size of the model, abundance of heavy elements, extraction temperature used in Section \ref{sec:methods}. For the model, the first letter D/S stands for double/single, while the second letter E/S stands for elliptical/spherical.}
    \label{tab:sample}
\end{table*}

The sample for this work is comprised of X-ray bright, low redshift galaxies, groups, and clusters that have been deeply observed with Chandra ($>$ 200\,ks) and which are disturbed by AGN feedback within the core regions.
We began with a sample of low redshift objects where X-ray cavities had previously been detected. 
The sample requirements for exposure time were dictated by the need for a sufficient number of high energy photons to conduct the analysis; although X-ray bright objects with less Chandra exposure time were explored, we were unable to discern any features in them using our method and thus did not include these objects in the sample. 
The sample is by no means complete, as we were limited to systems which already had long exposure times in the Chandra archive. 
The final sample of 15 objects can be found in Table~\ref{tab:sample}. Throughout this work, we often refer to the objects in this sample as a whole as ``gaseous halos'', meaning the gas-rich dark matter halos of systems ranging from galaxies to groups to rich clusters.

While most structures in our sample of objects are associated with AGN feedback, some of them are produced by mergers (e.g., sloshing of the gas). With X-arithmetic, we will probe the nature of \textit{all} structures imprinted on X-ray images.

\section{Data processing and image preparation} \label{sec:data}

For each object in our sample, we selected Chandra ACIS observations available in the archive (see Table~\ref{tab:obsids} for the summary). The data was reprocessed with \texttt{CIAO} version 4.15.2 with \texttt{CALDB} version 4.10.7 following the standard method described by \citet{Vikhlinin_2005}. {We also followed a standard approach to flare filtering, by first }extracting lightcurves of each obsID {and using the same binning as the ACIS blank-sky background files}, then removing time {bins} with count rates above 3$\sigma$ from the average count rate. The non-X-ray (instrumental) background image was created using re-scaled blank-sky images and accounting for readout background. Accounting for readout background involves modeling the contribution of the out-of-time events, a process discussed in more detail in \citet{Vikhlinin_2005}. For each observation and CCD, events in bad pixels were removed from the corresponding blank-sky background event file. The final event lists were merged, creating mosaic images of X-ray counts, a weighted exposure map, and the background. For each object, we produced images in soft and hard bands, the choice of which is discussed in Section~\ref{sec:methods}.

Once the final images were produced, the point sources were identified via the \texttt{CIAO} routine \texttt{wavdetect} and removed from the images. 
A model centered on the X-ray peak $(x_0, y_0)$ was then fit to the flux images to find the best-fit parameters.
The model options included spherically symmetric single and double $\beta$-models and elliptical single and double $\beta$-models. A single spherically symmetric $\beta$-model has the form
\begin{equation} \label{eq:sb_mod}
    S(R) = S(0) \left(1 + \frac{R^2}{r_c^2}\right) ^{-3\beta + 0.5}
\end{equation}
where $S(0)$ is the normalization, $r_c$ is the core radius, $R$ is the projected radius, and $\beta$ defines the slope.  For a spherical model, $R$ takes the form
\begin{equation}
    R = \left [(x - x_0)^2 + (y - y_0)^2 \right]^{1/2}.
\end{equation}
Elliptical models have the same form as Eq.~\ref{eq:sb_mod} except that the elliptical projected radius is given by
\begin{equation}\label{eq:ell_rad}
    R_e = \left [x_e^2 + \frac{y_e^2}{(1-\epsilon)^2}\right]^{1/2}
\end{equation}
where $\epsilon$ is the eccentricity and the coordinates are defined as 
$$x_e = (x - x_0) \cos(\theta) + (y - y_0) \sin(\theta) $$
$$ y_e = - (x - x_0) \sin(\theta) + (y - y_0) \cos(\theta)  $$
where $\theta$ is the angle of rotation of the ellipse. Double models are the sum of two $\beta$-models.

Visually inspecting the soft and hard band images and azimuthally-symmetric surface brightness profile shapes, we determined whether a (single/double) (spherical/elliptical) $\beta$-model was most suitable. Starting with reasonable initial values of the free parameters, a final best fit of the model to the image was found by optimizing the least square objective function using the Nelder-Mead algorithm included in the \texttt{scipy} optimization package.

{In the case of large-scale asymmetries, the best-fit model can be slightly modified to include these asymmetries into the model. Following \cite{Zhuravleva_2015}, this modified model is defined as $S_{\rm mod} = S\cdot G_{\rm \sigma}[I_{\rm X}/S]$, where $G_{\rm \sigma}$ denotes Gaussian smoothing with the window size $\sigma$ and $I_{\rm X}$ is the X-ray surface brightness image of a cluster. By adjusting $\sigma$, the model transitions from a smooth, symmetric $\beta$-model (large $\sigma$) to one that captures finer structures (small $\sigma$). In other words, decreasing $\sigma$ allows sharper features to be included in the model, leaving fewer structures in the residual image. This ``patching'' of the $\beta$-models, to varying degrees, was used to modify the model for every halo except the Phoenix Cluster.}

The resulting models are used to remove the large-scale gradients of surface brightness and the global trends in the temperature distribution, e.g. a strong radial dependence of temperature in cool-core clusters.
Dividing the soft- and hard-band images of a gaseous halo by corresponding models, we can obtain the residual images of gas perturbations in the ICM. Some of the structures, mainly the large-scale perturbations, may be captured and removed by the above procedure. However, small-scale perturbations should not be affected. Given all the choices for the unperturbed models, we only report results that are not sensitive to the choice of the model. 
Table~\ref{tab:sample} contains details of the chosen model for each object.

\section{Method} \label{sec:methods}

The method applied in this work, coined ``X-arithmetic'', has been developed and previously applied to the Virgo and Perseus clusters by \citet{churazov_2016}. It relies on assigning a particular ``effective equation of state'' to observed features, based on the relationship between amplitudes of temperature fluctuations $\delta T/T$ and gas density fluctuations $\delta n/n$:
\begin{equation}
    \frac{\delta T}{T} = \alpha \frac{\delta n}{n}
\end{equation}
where $\alpha = \gamma -1$, and $\gamma$ is the adiabatic index. X-arithmetic categorizes gas perturbations into three types: (i) weak shocks, sound waves, weakly-shocked gas (i.e., processes that do not change gas entropy, resulting in perturbations that appear ``adiabatic''), (ii) bubbles of relativistic plasma or X-ray cavities (can be interpreted as a variation of the thermal gas density, while the temperature remains the same, i.e. the perturbations are ``isothermal''), (iii) subsonic gas motions and gas cooling (i.e., pressure does not change, the perturbations will appear ``isobaric''). The corresponding values are then:
\begin{equation}
    \begin{aligned}
    \mathrm{adiabatic, }\ \alpha = 2/3,\  \gamma = 5/3 \\
    \mathrm{isobaric, }\ \alpha = -1,\ \gamma = 0 \\
    \mathrm{isothermal, }\ \alpha = 0,\ \gamma = 1.
    \end{aligned}
\end{equation}
By measuring a proportionality coefficient between density perturbations and temperature perturbations, the type of perturbation in the gaseous halo can be determined. 

Gas density and temperature have unique signatures in soft and hard energy bands of X-ray images. The X-ray emissivity per unit volume is
\begin{equation}
    f = n^2 \Lambda_B(T, Z)
\end{equation}
where $\Lambda_B(T, Z)$ is the X-ray emissivity, or cooling function, in a given energy band $B$, and is dependent on both temperature $T$ and abundance of heavy elements $Z$ (the dependencies are henceforth dropped for brevity, but implicit). For hot objects (with a temperature $\sim 2-10$\,keV), soft-band images are temperature-independent and reflect density perturbations, while the hard-band images are both density- and temperature-dependent \citep{Forman_2007}. Therefore, the three types of perturbations will be imprinted differently on the X-ray images in two bands. If we expand the X-ray volume emissivity by assuming that it is made up of an unperturbed and a perturbed component, such that $f = f_0 (1 + \frac{\delta f}{f_0})$, we can find the volume emissivity fluctuations $\delta f/f_0$ as a function of variables we can predict. For simplicity, we define $w = \frac{\delta f}{f_0}$. Finding $w$ requires that we assume (1) that $w$ can be decomposed into the sum of the three components:
\begin{equation}
   w = \sum_i w_i,
\end{equation}
where $i \in \{\mathrm{adiabatic,\ isobaric,\ isothermal\}}$, (2) that the amplitude of density fluctuations is small, (3) that the three types of perturbations describe all 
fluctuations (other types could be present, but these three will capture most structures), and (4) that second order terms can be neglected. It follows from these assumptions that
\begin{equation}\label{eq:perturbation}
    w_{B,i} = \left(\frac{\delta n}{n} \right)_{B,i} \left[2 + \alpha_i \displaystyle\frac{\mathrm d \ln \Lambda_B}{\mathrm d \ln T}\right].
\end{equation}

For a given observation, one can predict the ratio of X-ray surface brightness perturbations measured in two selected bands, hard $w_H$  and soft $w_S$, for pure isobaric, adiabatic and isothermal perturbations, using existing plasma emission codes. Conveniently, this ratio is independent of density perturbations $\delta n/n$:
\begin{equation}
    \frac{w_{H,i}}{w_{S,i}} = \frac{[2 + \alpha_i\left(\frac{\mathrm d\ln \Lambda_{H}}{\mathrm d \ln T} \right) | _{T_0}]}{[2 + \alpha_i\left(\frac{\mathrm d\ln \Lambda_{S}}{\mathrm d \ln T} \right) | _{T_0}]}
    \label{eq:flux_ratios}
\end{equation}
and thus, for a single type of perturbation ({each characterized by its own} $\alpha_i$), relies only on the value of the change in cooling function for each band, $\displaystyle\frac{\mathrm d\ln \Lambda_{S}}{\mathrm d \ln T}$ and $\displaystyle\frac{\mathrm d\ln \Lambda_{H}}{\mathrm d \ln T}$, at a chosen temperature $T_0$. X-ray residual images can, therefore, be linearly combined to suppress a given type of perturbation whose amplitude is predicted with this method. Visually inspecting the resulting images to determine which features are suppressed when a type of perturbation has been removed, one can classify perturbations based on their physical origin. A detailed description of how this classification is done can be found in the caption of Fig.~\ref{fig:a2052} for the example of A2052.

Although viewing these three final images and identifying features present in two out of three of them is the most reliable approach to identifying their nature, an alternative method that works best for high signal-to-noise objects and regions involves combining these maps into a single image, examples of which can be found in Appendix \ref{app:figures} and Fig.~\ref{fig:3color}.

\subsection{Calculating the proportionality coefficient}

\begin{figure}[ht!]
    \centering
    \includegraphics[width=0.95\linewidth]{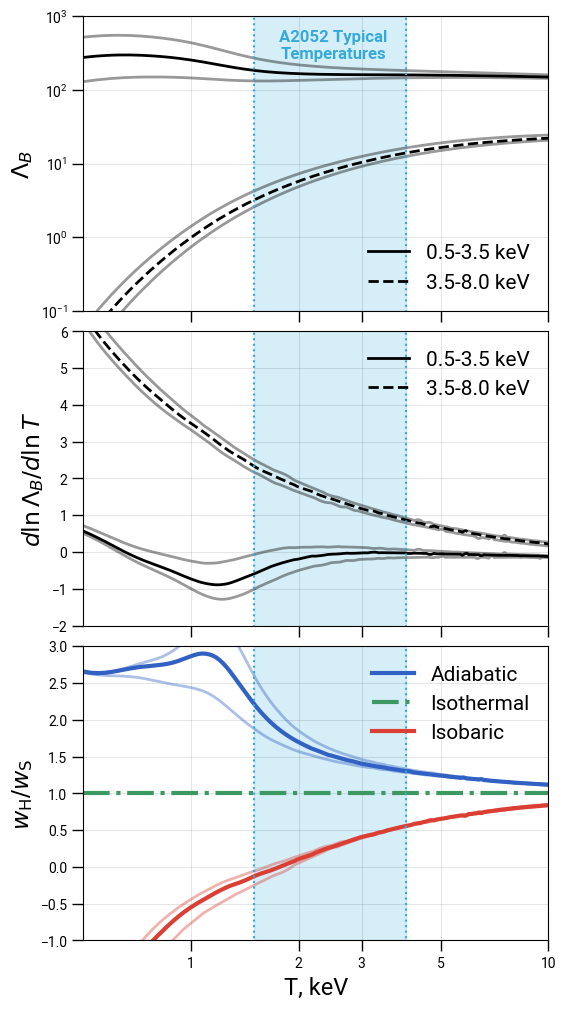}
    \caption{Top: $\Lambda_B$ for the soft (solid line) and hard (dashed line) bands of A2052 calculated using XSPEC and APEC. Middle: $\displaystyle\frac{\mathrm d\ln \Lambda_B}{\mathrm d \ln T}$ as a function of temperature for the soft (solid line) and hard (dashed line) bands of A2052. Bottom: Ratio of fluxes in the hard to soft band for different types of perturbations vs. gas temperature in A2052 obtained using equation \ref{eq:flux_ratios}. In this cluster, pure adiabatic (isobaric) perturbations are stronger in the hard (soft) band. The lighter lines (gray in top two panels, blue and red in third panel) show how the ratios vary for a reasonable range (0.2 to 1.0 solar) of abundances.}
    \label{fig:a2052_curves}
\end{figure}

\begin{figure}[ht!]
    \centering
    \includegraphics[width=0.95\linewidth]{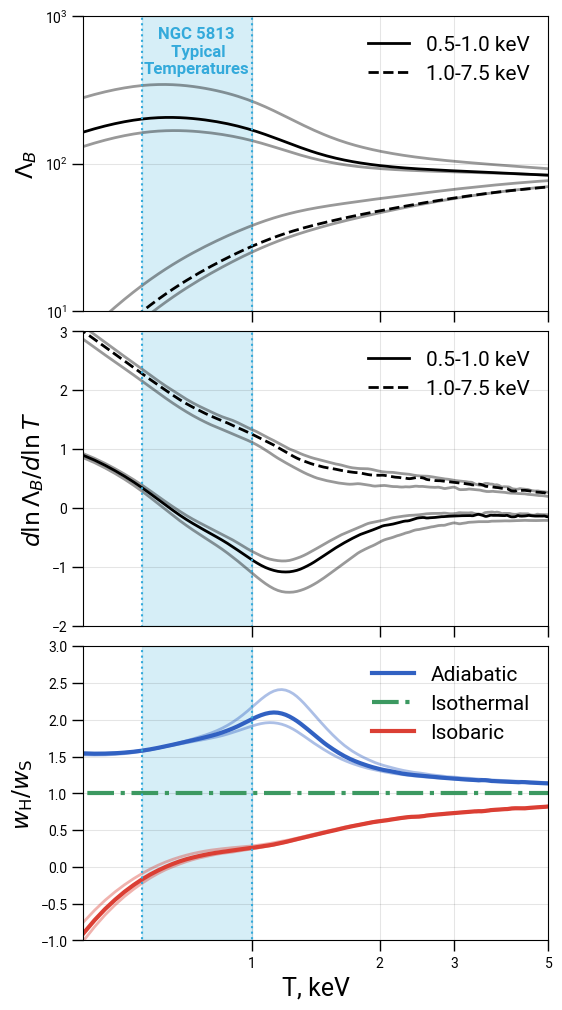}
    \caption{Same as caption in Fig. \ref{fig:a2052_curves} but for NGC 5813. The range of abundances used is 0.3 to 0.7 solar.}
    \label{fig:ngc5813_curves}
\end{figure}

As examples of our procedure, we will use A2052 (a cluster with $M_{500}$ mass $2.49 \times 10^{14}\ \mathrm{M}_\odot$, \citet{Piffaretti_2011}) and NGC 5813 (a galaxy group with mass $3.9 \times 10^{11}\ \mathrm{M}_\odot$, \citet{Cappellari_2013}) to illustrate the steps in calculating the proportionality coefficient. In practice, this process involves computing the response files (ARF and RMF) for an observation, then simulating spectra using XSPEC \citep{Arnaud_1996} and AtomDB for a given abundance $Z$, redshift $z$, and hydrogen column density $n_{\rm H}$ to determine how X-ray emissivity varies with temperature for two bands. The emissivity curves $\Lambda_{\rm B}$ for two bands, simulated using this method, can be seen for A2052 and NGC 5813 in the top panels of Fig.~\ref{fig:a2052_curves} and Fig.~\ref{fig:ngc5813_curves}, respectively. Within the ranges of typical temperatures for each object, the soft band tends to have higher values than the hard band by an order of magnitude. It is then straightforward to derive $\frac{\mathrm d\ln \Lambda_{\rm B}}{\mathrm d \ln T}$ from these curves, shown in the middle panels. The hard band typically varies more than the soft band, so $\frac{\mathrm d\ln \Lambda_{\rm H}}{\mathrm d \ln T}$ is usually a factor $\sim 2$ larger than $\frac{\mathrm d\ln \Lambda_{\rm {\rm S}}}{\mathrm d \ln T}$.

The ratio between two bands $w_{\rm H,i}/w_{\rm S,i}$ can then be calculated for each type of perturbation using equation~\ref{eq:flux_ratios}. The bottom panels of Fig.~\ref{fig:a2052_curves} and Fig.~\ref{fig:ngc5813_curves} show these predicted values for A2052 and NGC 5813, respectively.  One can see that for the chosen soft and hard bands, isobaric (adiabatic) perturbations have higher amplitudes at softer (harder) energies than at harder (softer) energies. As {can be} expected, isothermal perturbations remain the same across soft and hard energy bands. For A2052, adiabatic values are between 1.25 and 2.5 for the range of typical temperatures of the object, while isobaric values range from -0.25 to 0.5 and isothermal values stay consistently at 1.0. For NGC 5813, the adiabatic coefficient varies between 1.5 and 2.25 for typical temperatures, the isobaric coefficient between -0.25 and 0.25, and the isothermal remains at 1.0. The range of reasonable abundances for each object (0.2 to 1.0 solar for A2052 and 0.3 to 0.7 solar for NGC 5813), the effect of which is shown via lighter lines, is based on previous spectral studies, and typically leads to a $\leq 15\%$ variation in the calculated ratio, with the variance larger for the adiabatic perturbations than the isobaric due to the differing coefficients.

The choice of extraction temperature for the coefficient is a systematic uncertainty of the method, as all objects have a range of temperatures within their central region. For A2052, for example, $\frac{\mathrm d\ln \Lambda_B}{\mathrm d \ln T}$ varies by a factor of $\sim 2$ for the hard band and $< 1$, close to temperature independent, for the soft band (Fig.~\ref{fig:a2052_curves}) within its temperature range. This trend differs for cooler systems like NGC 5813 (Fig.~\ref{fig:ngc5813_curves}), where the soft band varies as much or more than the hard band. In most systems, the values for both bands vary by no more than a factor of 2 within the temperature range. Additional systematics of this step include the observation cycle of the ObsIDs, since only one set of response files can be used to predict the proportionality coefficient, and in the simulation of spectra, the abundance of heavy elements and gaseous halo temperature. Dealing with superposition along the line of sight is discussed in Section~\ref{sec:proj}. We only report those features which are stable against all systematics, including the choice of unperturbed model discussed in Section~\ref{sec:data}. Default choices for each object in the calculation of the coefficient are collected in Table~\ref{tab:sample}.

\subsection{Choosing energy bands}

Although the soft X-ray band does trace the density fluctuations in objects of typical cluster temperatures (see Fig.~\ref{fig:a2052_curves} top panel), temperature independence is not required for the analysis as long as the calculation of $w_{\rm H,i}/w_{\rm S,i}$ is tailored to a given observation. 
X-ray analysis is limited by hard band photon statistics, so, in practice, the choice of the soft and hard energy bands was based on (a) assigning enough counts to the hard band to be able to identify features
and (b) the largest possible difference between pure isobaric and adiabatic predictions (see the difference between red and blue curves within the blue regions in Fig.~\ref{fig:a2052_curves} bottom panel). Our choices of soft and hard band for each object are collected in Table~\ref{tab:sample}.

\subsection{Projection effects}\label{sec:proj}

The final step before combining the images is to account for projection effects. The observed X-ray emissivity is the projection of the volume emissivity $f$ along the line of sight. The ``reprojection'' coefficient at each line of sight \cite[see][]{Churazov_2012, Zhuravleva_2012} depends on the surface brightness profile, which is approximated with a model. The different best-fitting models of the soft and hard band images lead to different projection effects in the images of perturbations in these two bands. To correct for this effect, we calculate analytically a projection correction factor $X(R)$ that depends on the parameters of the best-fit model, following \cite{churazov_2016}. To simplify this calculation, the best-fit single $\beta$-model, either spherical or elliptical, was used in place of the more complex options introduced in Sec.~\ref{sec:data}. Fig.~\ref{fig:SBprof} in Appendix \ref{app:figures} shows three examples of surface brightness profiles with their best-fit default models (listed in Table~\ref{tab:sample}) and their best-fit single models, as an example of how closely the two models resemble one another and to motivate this simplification.

For the calculation of the projection correction factor $X(R)$, we begin with the understanding that the volume emissivity along the line of sight is made up of an unperturbed and a perturbed gas distribution, as assumed earlier in Section~\ref{sec:methods}. Therefore, the observed image is
\begin{multline}\label{eq:sb_def}
    S(x,y) = \int f \mathrm dz = \int n^2\Lambda_B \mathrm dz \\ = \int n_0^2 \Lambda_B\mathrm dz  + \int n_0^2 \Lambda_B \left[2 + \alpha \frac{ln \Lambda_B}{ln T}\right] \frac{\delta n}{n} \mathrm dz \\
    = S_0(x,y) + \int n_0^2 \Lambda_B \left[2 + \alpha \frac{ln \Lambda_B}{ln T}\right] \frac{\delta n}{n} \mathrm dz
\end{multline}
where $n_0$ is the undisturbed density distribution and $S_0$ is the undisturbed surface brightness.

We then assume that the perturbation is located at $z = z_0$ and has a spatial extent $\Delta z$. Rearranging, this allows us to write:
\begin{equation}
    \left(\frac{S}{S_0} -1\right) \frac{S_0}{n_0^2(r)\Lambda_B} = \left[2 + \alpha \frac{ln \Lambda_B}{ln T}\right] \frac{\delta n}{n} \Delta z.
\end{equation}
Here, we have used $r = \sqrt{x^2 + y^2 +z_0^2}$. $ \left(\frac{S}{S_0} -1\right)$ is, by definition, the residual/perturbation image so $\frac{S_0}{n_0^2(r)\Lambda_B}$ is the position-dependent correction factor that we want to solve for.
Starting with the numerator, we take the definition of surface brightness (the first line of Eq.~\ref{eq:sb_def}), replace $n_0^2\Lambda_B$ with our $\beta$-model definition, and use the relationship $r^2 = R^2 + z^2$. This yields 
\begin{multline}
    S_0 = 2 \int_0^\infty \frac{S(0)}{\left(1 + (\frac{r}{r_c})^2 \right)^{3\beta}} dz  \\ = \frac{S(0)(1 + (R/r_c)^2)^{0.5-3 \beta} r_c \sqrt{\pi} \Gamma(3 \beta-0.5)}{ \Gamma(3 \beta)}.
\end{multline}
To simplify the denominator $n_0^2(r)\Lambda_B$, we assume that the observed perturbation is located at the midline of the gaseous halo $z=0$ such that
\begin{equation}
    n_0^2(R) \Lambda = \frac{S(0)}{ \left(1 + (R/r_c)^2 \right)^{3 \beta}.}
\end{equation}
The correction factor $X(R)$ is then 
\begin{equation}
    X(R) = \frac{\sqrt{\pi}(1 + \left( \frac{R}{r_c}\right)^2)^{0.5}r_c\Gamma(3\beta - 0.5)}{\Gamma(3\beta).}
\end{equation}
where $r_c$ and $\beta$ are the values from the best-fit single model. For the case of an elliptical $\beta$-model, the derivation goes similarly, but using the elliptical projected radius $R_e$ introduced in Eq.~\ref{eq:ell_rad} in place of $R$.
We then scale the hard-band residual image by the ratio of projection corrections of the soft band to the hard band, eliminating the position-dependence so that, effectively, the hard-band residual image is brought to the soft-band image (i.e., the projection effects are the same in both images). 
The hard and soft band images can then be combined such that a single type of perturbation is eliminated in each of three final images. Final images were adaptively smoothed to allow for clearer feature identification. {This smoothing can result in artifacts at higher radii, where photon counts are low. Features were only identified if they were prominent and persistent regardless of smoothing.}

\section{Results} \label{sec:results}

\subsection{Cluster: A2052}

\begin{figure*}
    \centering
    \includegraphics[width=\linewidth]{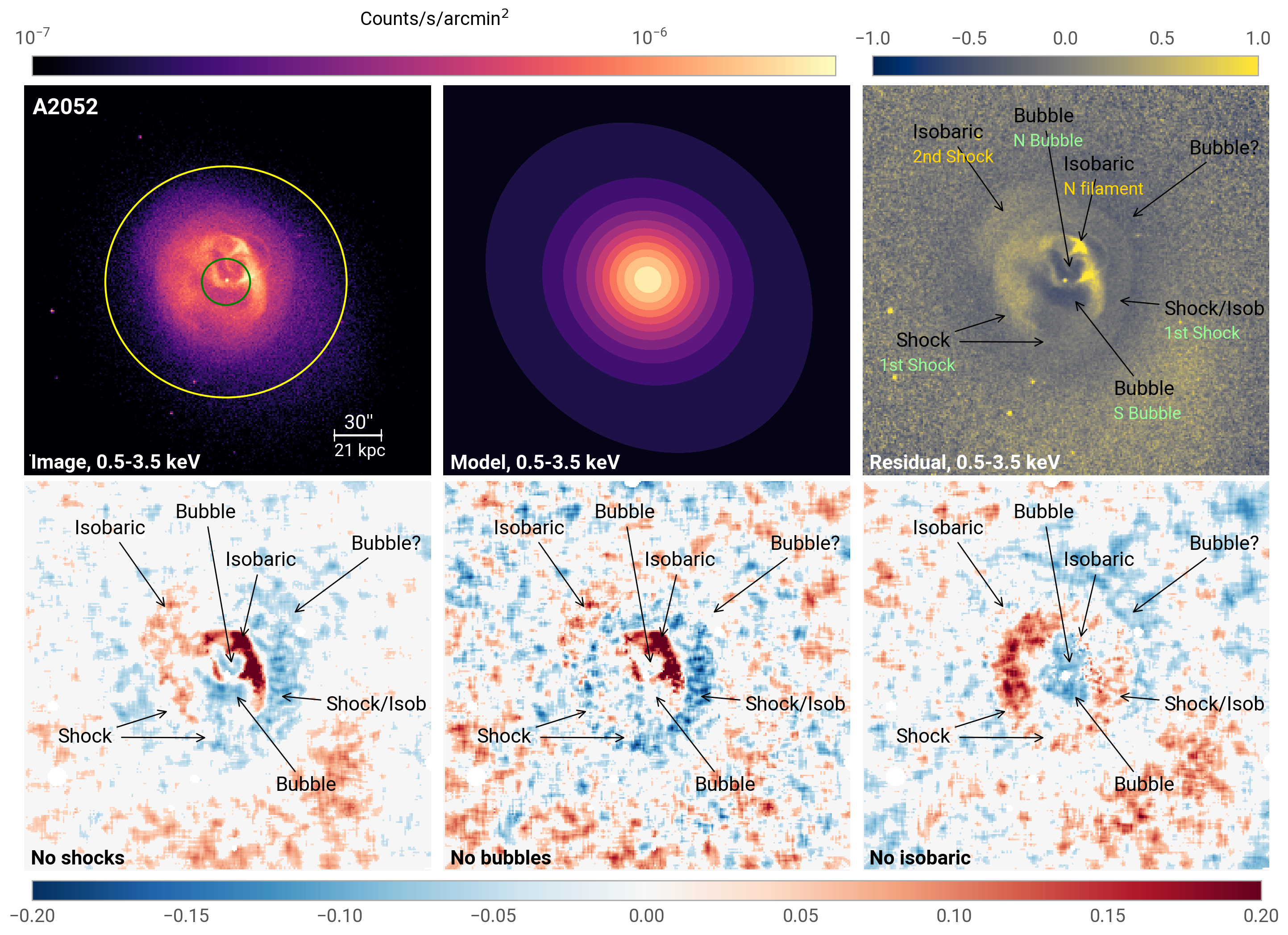}
    \caption{Application of X-arithmetic to A2052. \textit{Upper left:} Soft band (0.5-3.5\,keV) Chandra image. Circles provide an easy comparison of size scales: cyan circles have a 5\,kpc radius, green circles -- a 10\,kpc radius, yellow -- a 50\,kpc radius, and red -- a 200\,kpc radius. Only those circles which are reasonably visible are included in the soft band image of each system. \textit{Upper middle:} Global model of X-ray surface brightness distribution in the soft band (0.5-3.5\,keV). \textit{Upper right:} Soft band (0.5-3.5\,keV) residual image (the image divided by the model). Point sources are shown, but were removed to perform the analysis. Black labels point to features we identified via the X-arithmetic method, while green and gold text shows previous identification of these features \citep{Blanton_2011}. The green text identifies features which were either confirmed by or did not conflict with the X-arithmetic images, while the gold text highlights features which differed from or were further elucidated by our method. \textit{Bottom row:} The soft and hard band residual images are combined to show, from left to right, adaptively smoothed maps which exclude, in turn, perturbations caused by (1) weak shocks and sound waves (adiabatic fluctuations), (2) bubbles (isothermal), and (3) subsonic motions and/or gas cooling (isobaric). A central isobaric feature, originally called the northern filament, is evident in the first two images but absent (close to zero, i.e. white) in the third. Bubbles, as expected, are absent in the central image,  but can be identified south of the AGN on the right and left images. A shock front and shock-heated gas, originally identified as the first shock, which appear as a ring around the central regions, are absent in the first image. The SW part of the ``shock'' structure is a mix of isobaric and shock perturbations. The NE feature earlier tentatively identified as the second shock has an isobaric nature and is likely associated with sloshing of the gas.
    }
    \label{fig:a2052}
\end{figure*}

A2052 is a cool-core cluster with a powerful radio source associated with its central cD galaxy that has both a compact core and a diffuse, extended component \citep{Zhao_1993}. Its long Chandra exposures have been analyzed in, e.g., \citet{Blanton_2009, Blanton_2011}. These works identify inner and outer cavities around the central AGN coincident with radio synchrotron emission. They also find a nearly circular shock front of Mach number $\sim 1.2$, and an additional potential shock to the northeast. A large-scale spiral feature is likely evidence of gas sloshing. For earlier-identified structures, see green and gold notations in the top right panel of Fig.~\ref{fig:a2052}. Gold text highlights those elements from other authors that X-arithmetic either identified differently or further clarified, while green identifies features confirmed or consistent with X-arithmetic analysis.

 Applying X-arithmetic to A2052, the most significant structures include the central bubbles, both north and south of the central AGN, and the isobaric structure to the northwest of the bubbles (Figure~\ref{fig:a2052}, bottom panels). This feature was previously identified as the ``northern filament'' and is spatially aligned with H$\alpha$ contours \citep{McDonald_2010}, indicating the likely presence of cooling gas. The prominent ring-like structure around the bubbles, called the first shock in other works, appears along most of its arc as a mix of shock-heated gas (adiabatic perturbations) and isobaric-like perturbations.
 A mix of perturbations could also be two independent structures that overlap in projection, leading to an ``intermediate'' equation of state that is not fully removed from either the ``no shocks'' or ``no isobaric'' image.
 To the northeast, in the region tentatively identified as a second shock, this method finds an effective equation of state that better matches an isobaric feature. This could be related to gas sloshing discussed in \citet{Blanton_2011}, the evidence for which is a cool, high abundance spiral in the surface brightness at distances larger than shown in Figure\,\ref{fig:a2052}. An additional feature that was not previously studied, but which is stable against all systematics, lies to the northwest at a similar radius as the isobaric feature, and appears to be a bubble. This is marked with a question mark in the figure due to the high noise, but could be an interesting target for deeper X-ray and high-sensitivity, low-frequency radio observations in the future. 

 At even larger radii, spiral features (i.e., the bright region to the south) are present in the residual images and the X-arithmetic images. However, these features are too strongly affected by noise to identify their nature robustly.

\subsection{Cluster: Centaurus}

The Centaurus cluster of galaxies is one of the X-ray brightest and nearest clusters. Its Chandra data were analyzed in depth in \citet{Sanders_2016}. This work uncovered central cavities surrounded by shocks, prominent X-ray plumes, a hook-like structure, and evidence of large-scale gas sloshing.

\begin{figure*}
    \centering
    \includegraphics[width=0.85\linewidth]{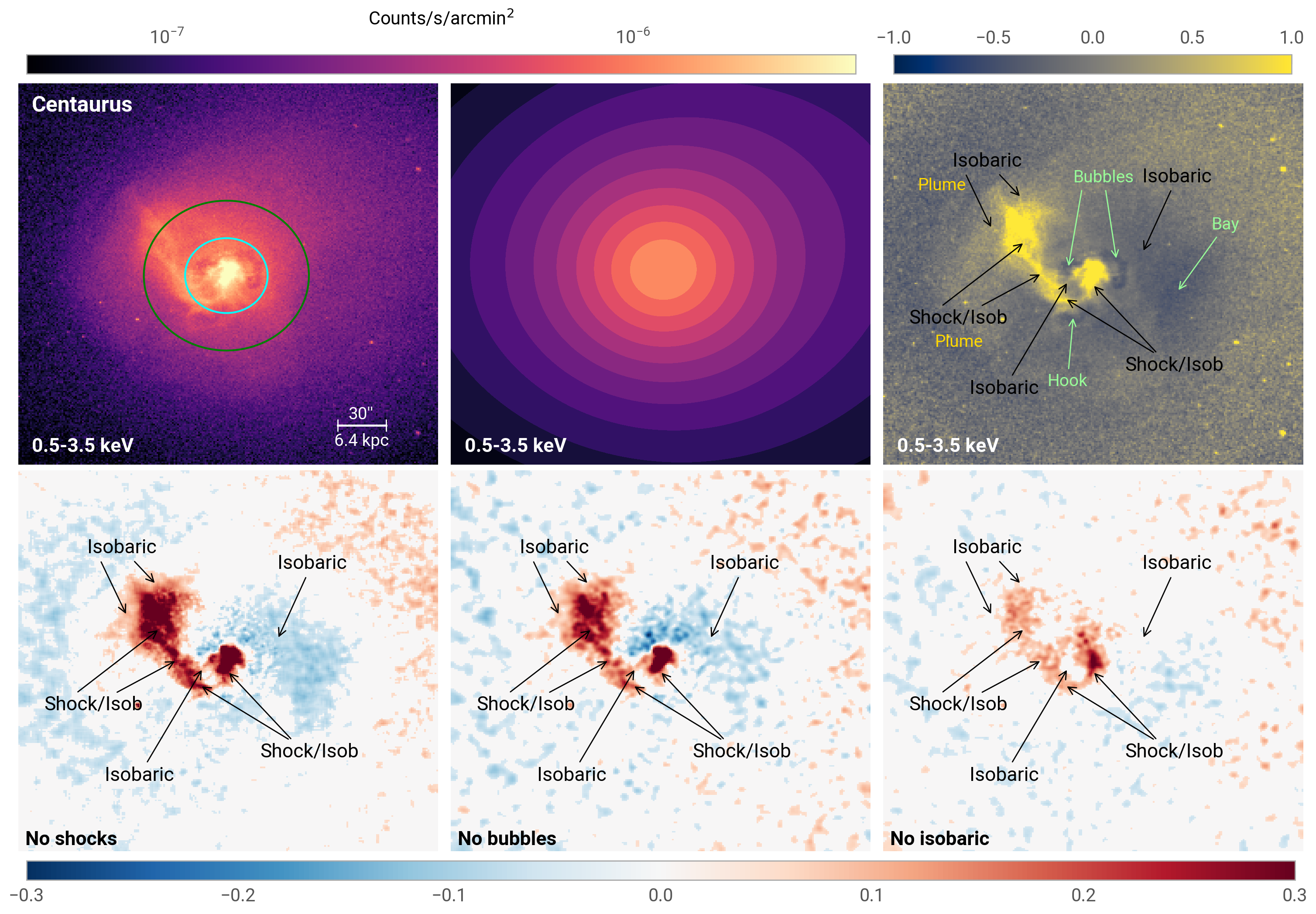}
    \includegraphics[width=0.85\linewidth]{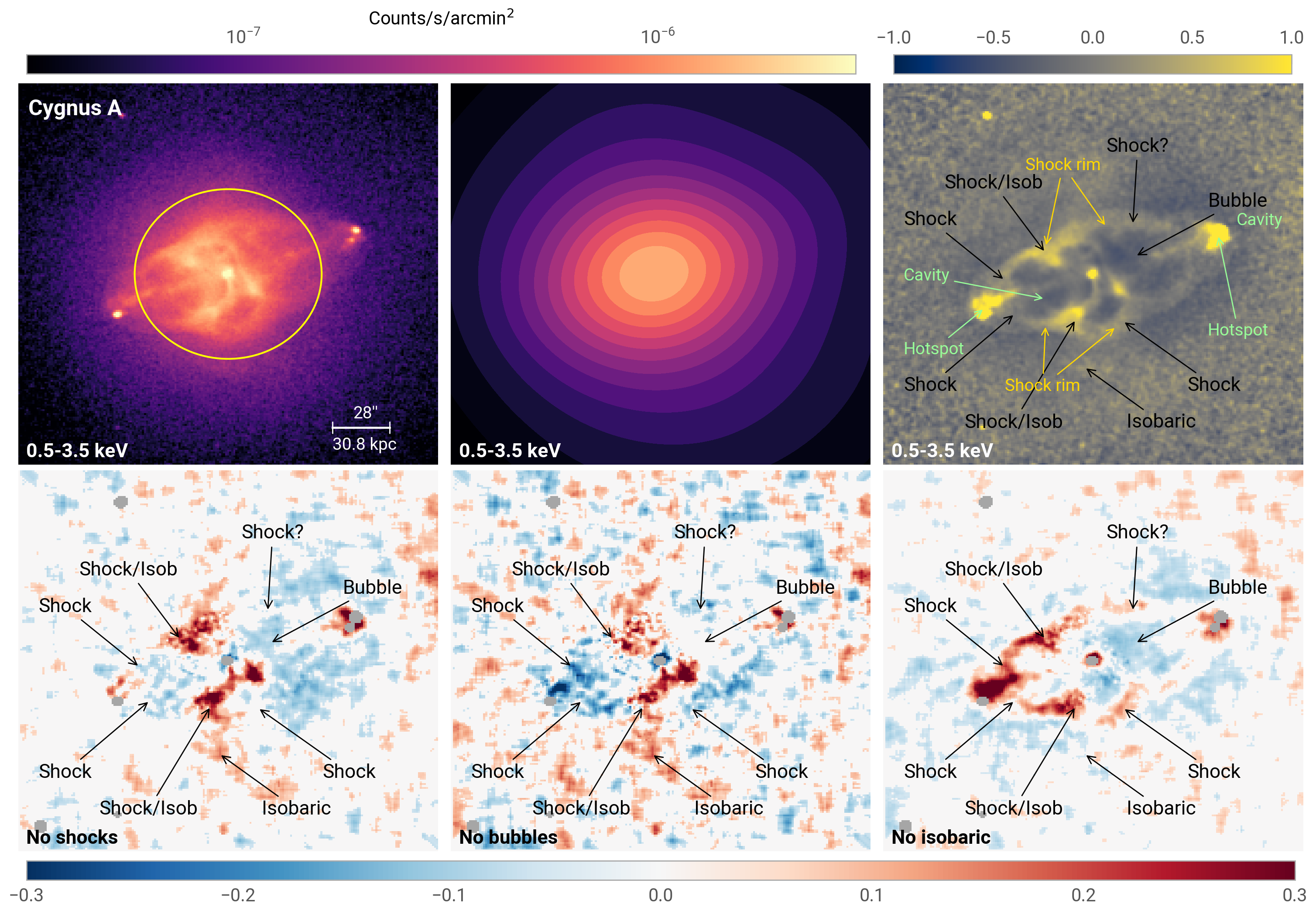}
    \caption{Application of X-arithmetic to the Centaurus cluster (top) and Cygnus A (bottom). Panel order and notations are the same as in Figure~\ref{fig:a2052}. 
    In Centaurus, the plume is found to be a mix of shocked and isobaric gas at its core, surrounded by a diffuse isobaric component. The bay appears isobaric in nature, and another isobaric filament connects X-ray enhancements on either side of the eastern bubble.
    In Cygnus A, the so-called shock rim is comprised of shocked gas to its eastern side and a mix of shocked and isobaric emission at its brightest points. It surrounds a cavity, which can be confirmed on the western half. A new isobaric feature is located to the south.
    }
    \label{fig:centaurus}
\end{figure*}

Images were produced with temperatures and abundances tailored to each of three regions within the cluster images; for the center, 1.2\,keV and 0.5\,Z$_{\odot}$, for the left, 3.7\,keV and 0.7\,Z$_{\odot}$, and for the right, 2.5\,keV and 1.5\,Z$_{\odot}$. Although each version of the final images made certain features more or less apparent, only those features stable to systematics are labeled in the upper panels of Figure~\ref{fig:centaurus}. In the bottom row of this figure, images using 0.5\,Z$_{\odot}$ and 1.2\,keV are shown. The images tailored to the central region seem to work best overall, since at higher extraction temperatures the final images become noisier due to less separation between the isobaric and adiabatic coefficients.

Examining the features, the plume is a mix of shocked and isobaric perturbations at its core. The more extended emission of the plume, at larger radii, is purely isobaric. Bubbles and shocks were previously identified in the central region, but X-arithmetic identifies mixed shocks and isobaric perturbations. No bubbles can be identified, likely because they are too small and the data too noisy for identification with the current exposures. There is, however, a delicate isobaric filament that stretches between two prominent regions of mixed shocks and isobaric emission. An extended region to the right (west) near the bay appears isobaric, and may be related to larger scale gas sloshing present in the cluster.

On scales larger than shown in Figure~\ref{fig:centaurus}, large-scale spirals are evident in the cluster. For many choices of extraction temperature, they appear to be isobaric in nature, which would match their interpretation as being due to gas sloshing. However, these features are in noisier regions and not stable against all systematics.

\subsection{Cluster: Cygnus A}

Cygnus A is the prototypical FR II radio galaxy. As such, and as one of the brightest extragalactic sources in the X-ray sky, it has been extensively studied. Previous works with a focus on the Chandra data include \citet{Wilson_2000, Young_2002, Smith_2002, Balucinska_2005, Wilson_2006, Steenbrugge_2008, Snios_2018}. The Cygnus A galaxy lies at the center of an identically named cool-core galaxy cluster and exhibits powerful collimated radio jets, a cocoon shock, and several X-ray hot spots. Radio emission is coincident with the eastern and western cavities that are internal to the cocoon shock \citep{Snios_2018}.

The application of X-arithmetic to Cygnus A is complicated by the large-scale asymmetry of the cluster. 
The western half of the ``cocoon'' structure is dimmer than the eastern, a feature which is captured by the default model.
In our images, seen in the bottom row of Fig.~\ref{fig:centaurus}, we can consistently confirm much of the eastern half of the previously identified cocoon shock as a shock, with the brighter nodes $\sim 30$\,kpc from the AGN appearing as a mixture of shocked and isobaric gas. 
The western half appears as a shock with the default model; however, it appears to be a mixture of shocked and isobaric emission with the use of a global model with a larger patching radius (i.e., 70'' rather than 30'') that does not remove as much of the large-scale asymmetry. 
To remain true to our commitment to only identify structures which are unchanged by systematics, this extent of the cocoon is labeled with a question mark.
In the western half of the cluster, we can confirm an X-ray cavity in a region coincident with a radio bubble. 
An extension to the south appears isobaric.

\subsection{Cluster: M87}
M87, the central galaxy of the Virgo Cluster, is one of the most well-studied AGN feedback systems. Its proximity allows for resolution of its jet and of perturbations in its $\leq 2$\,keV ICM. Numerous studies \citep[a recent sample includes][]{Million_2010, Simionescu_2017, McCall_2024} have explored the galaxy cluster in the X-ray band. Its Chandra data alone, focused on the regions most affected by AGN feedback, has been the feature of many works, including \citet{Churazov_2001, Young_2002b, Forman_2007} and references therein. \citet{Forman_2007} identified eastern and southwestern filamentary arms, which they attribute to potentially arising from buoyant bubbles. The authors also point to a shock at 13\,kpc, a high pressure region around the AGN and jet, and an outer ring of enhanced X-ray emission. A work applying X-arithmetic to M87 \citep{churazov_2016, Arévalo_2016} identified the arms as isobaric, a weak shock both at 13 kpc and also within an inner region, and bubbles to the east of the jet and also just outside of the eastern arms. 

\begin{figure*}
    \centering
    \includegraphics[width=0.85\linewidth]{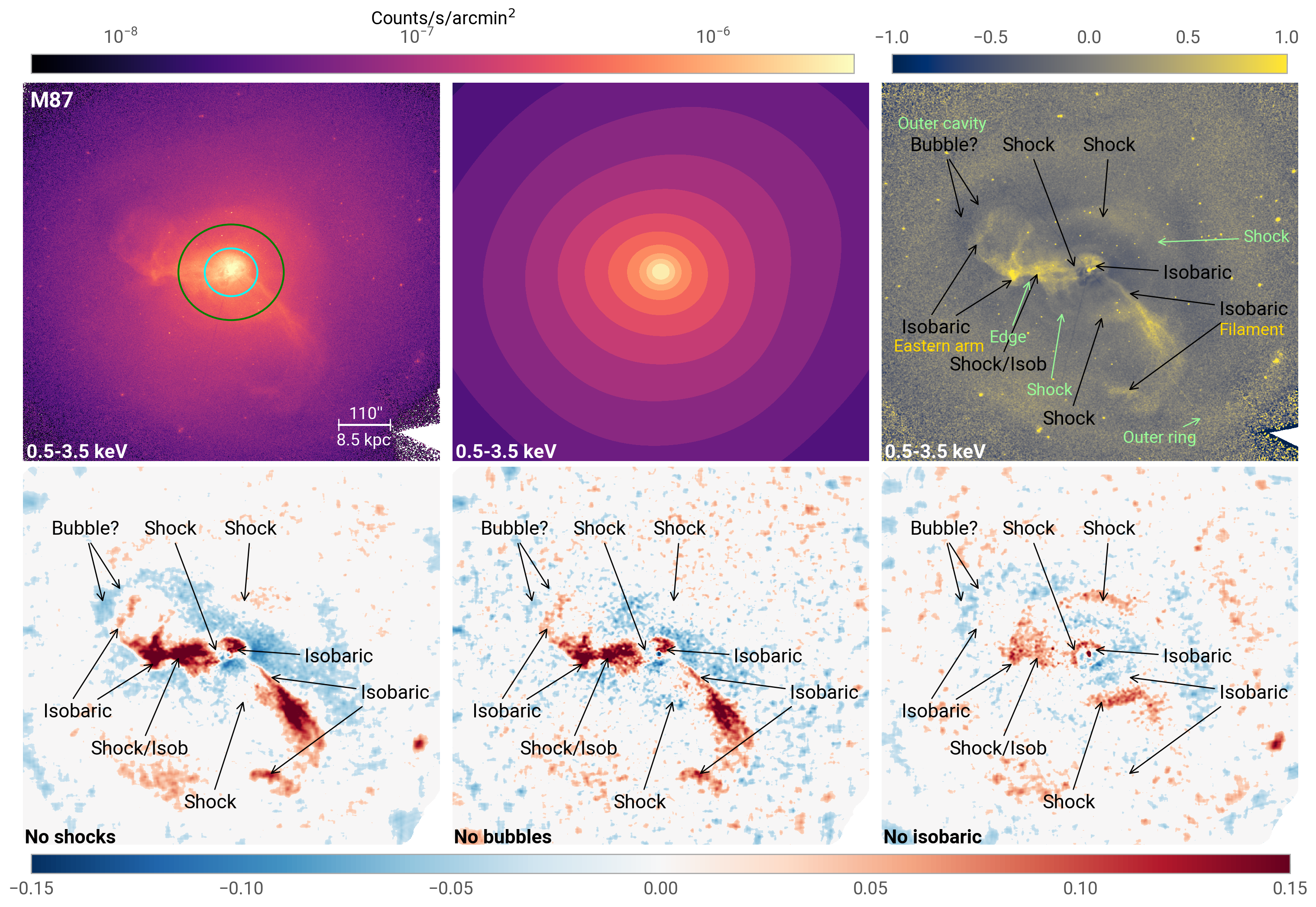}
    \includegraphics[width=0.85\linewidth]{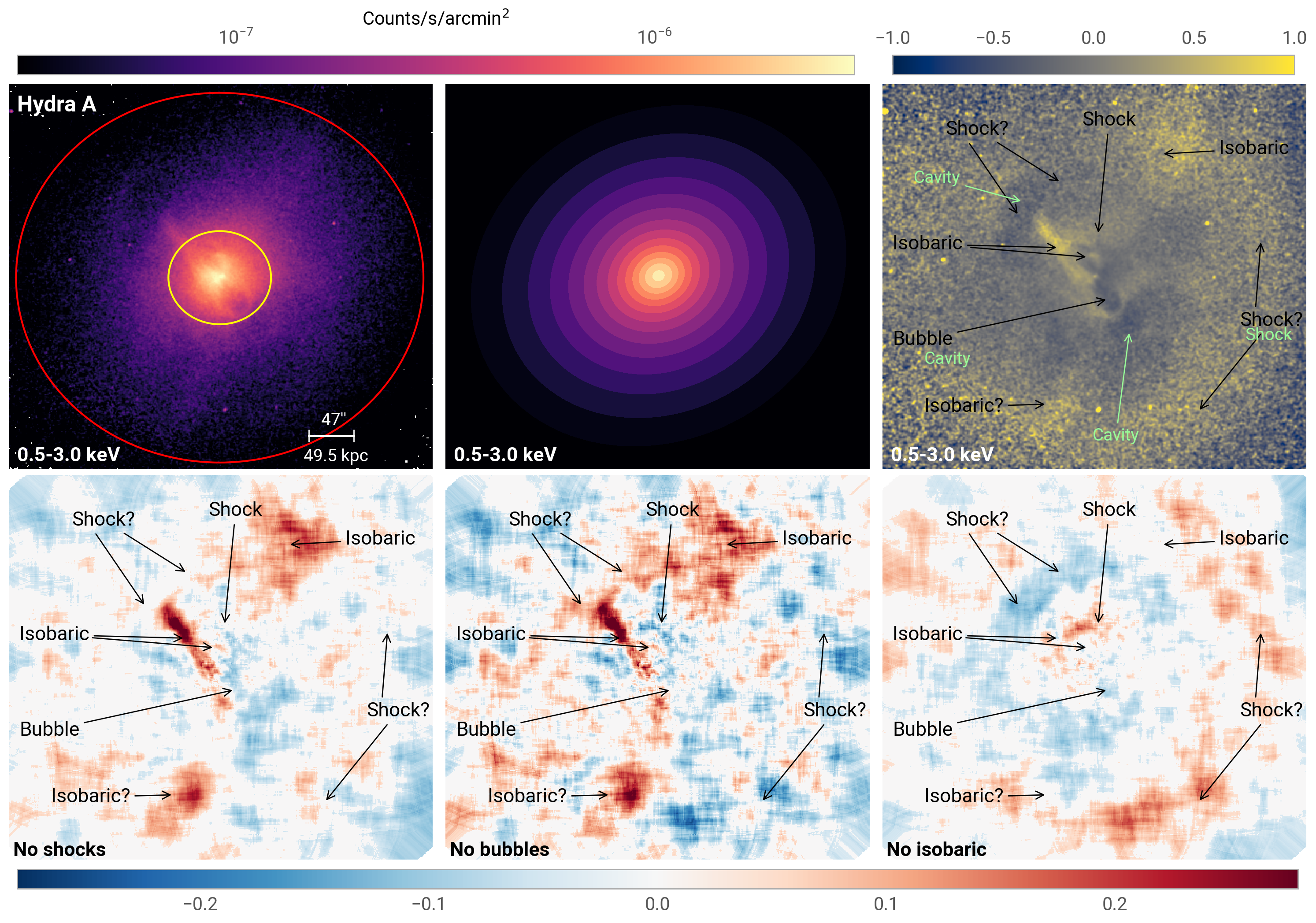}
    \caption{Application of X-arithmetic to M87/ Virgo cluster (top) and Hydra A (bottom). The panel order and notations are the same as in Fig.~\ref{fig:a2052}.
    In M87, the eastern arm is comprised of a shock and isobaric mix at its core, with a diffuse isobaric component at its edges. The southwestern arm, or filament, appears isobaric. We confirm multiple regions of shocked gas.
    In Hydra A, we confirm the presence of a southern cavity and potential shocked gas to the southwest. We also identify several isobaric features and a shock to the north of the northern cavity.
    }
    \label{fig:m87}
\end{figure*}

In this updated analysis, seen in the top half of Fig.~\ref{fig:m87}, we also identify the isobaric arms. Similarly to the Centaurus X-arithmetic results for the plume, the brighter central region of the eastern arm is a mix of shocks and isobaric, but exclusively isobaric at larger radii. Portions of two shocks are clearly confirmed both in the inner region and farther out at 13\,kpc. The outer cavity is identifiable as a bubble when lower extraction temperatures, best suited to the inner regions, are used.

\subsection{Cluster: Hydra A}
The Hydra A cluster of galaxies is known for its prominent X-ray cavities, coincident with radio lobes \citep{Wise_2007, Simionescu2009, mcnamara_2000}. A large-scale shock surrounds the cavities \citep{Simionescu2009}.

With X-arithmetic (Fig.~\ref{fig:m87}), we identify an isobaric structure pointing northeast between two generations of cavities, with a shocked region directly to the west. Only a southern cavity is visible. A candidate shock front lies just beyond the isobaric arm to the north, in the southern part of a region believed in previous works to be a cavity. This feature is temperature-dependent. A mix of perturbations surrounds the cluster center to the southwest, including a candidate isobaric enhancement and a candidate shock front that are difficult to identify due to low statistics. However, a distinct isobaric enhancement appears in the northwest region at the same radius, robust against different choices of temperature and abundance. 

\subsection{Cluster: Phoenix}
The Phoenix cluster is known as an outlier in the world of cool-core clusters due to its high redshift (z = 0.596), high X-ray luminosity, and massive central starburst ($\sim 610\,\mathrm{M}_{\odot}\,\mathrm{yr}^{-1}$; \cite{McDonald_2015, Tozzi_2015}). It contains a quasar at its center, and also has powerful radio jets; its power is estimated to be half quasar-mode and half radio-mode \citep{McDonald_2019}. Two X-ray cavities have been discovered to coincide with the radio jets \citep{McDonald_2019}. \citet{McDonald_2015} also proposed the existence of over-dense regions around the central cavities and two additional ghost cavities at larger radii. A spiraling cool gas feature was first reported with low significance by \citet{McDonald_2015} and confirmed to be coincident with a radio mini-halo by \citet{Raja_2020}.

\begin{figure*}
    \centering
    \includegraphics[width=0.85\linewidth]{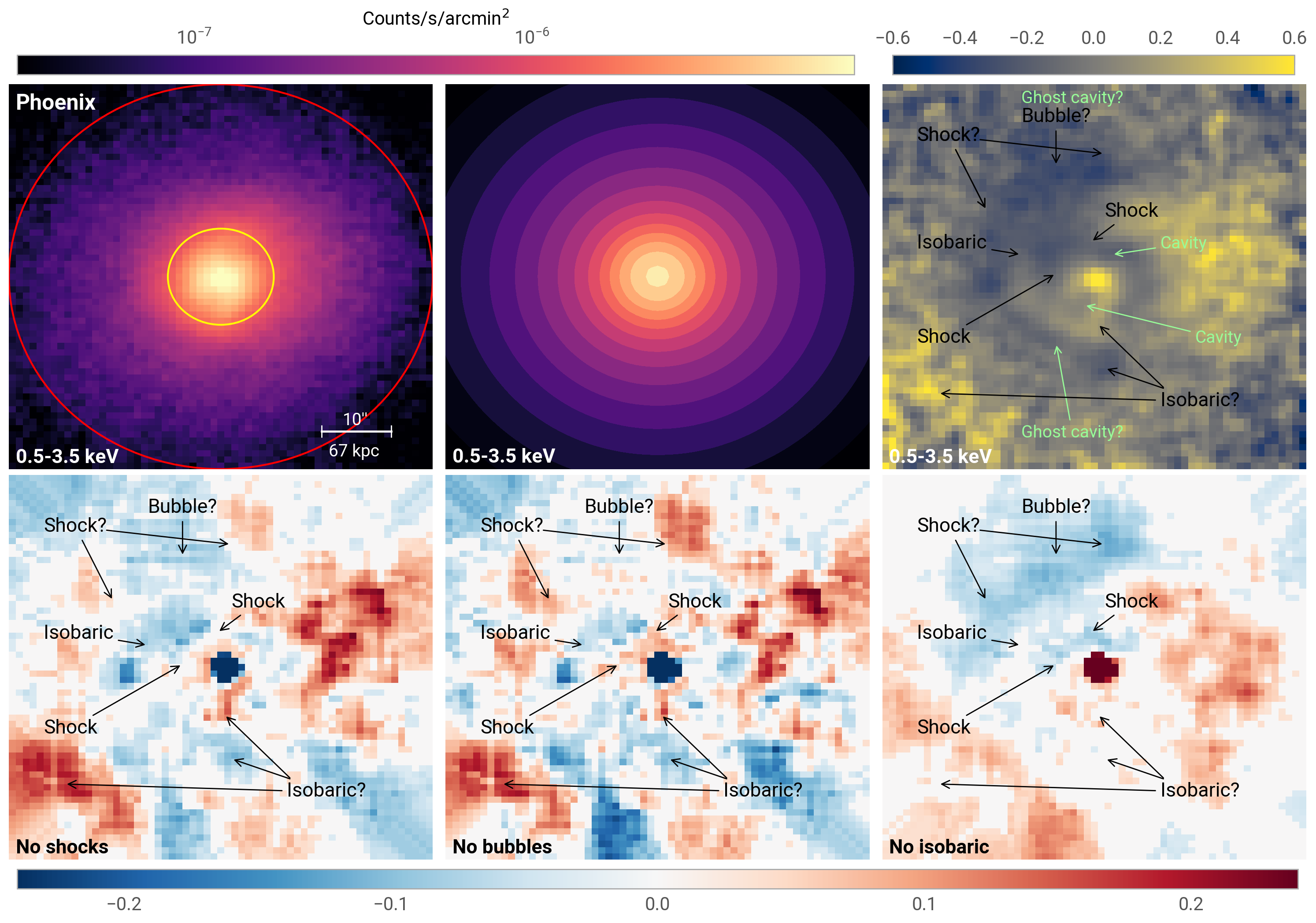}
     \includegraphics[width=0.85\linewidth]{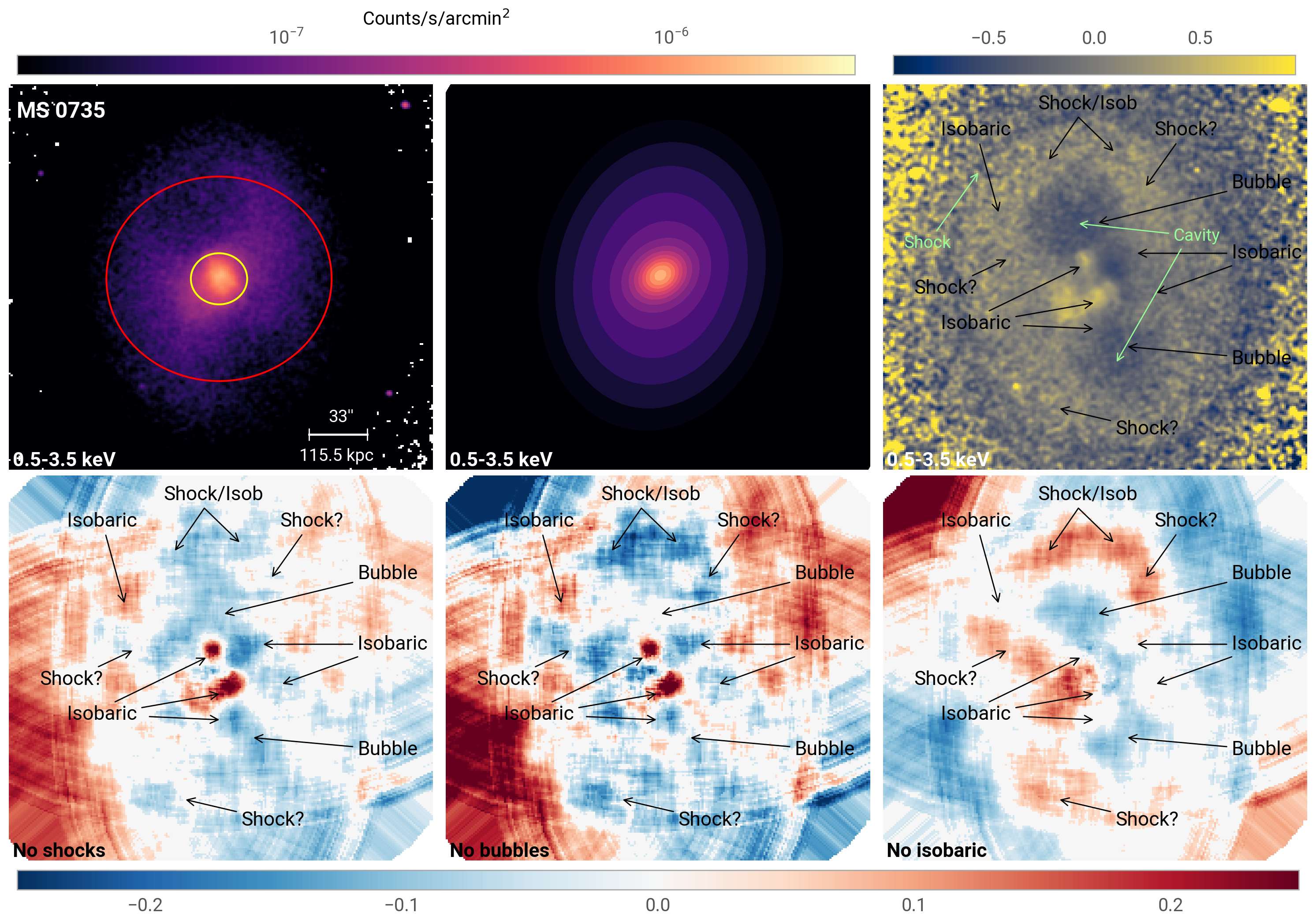}
    \caption{Application of X-arithmetic to Phoenix (top) and MS 0735 (bottom). The panel order and notations are the same as in Fig.~\ref{fig:a2052}.
    In Phoenix we identify two shocked regions within 50\,kpc. Several candidate isobaric features can be identified outside this inner region, in addition to a candidate bubble and candidate shocked gas to the north.
    In MS 0735 we confirm the two prominent bubbles. The central emission (within $\sim$100\,kpc) appears isobaric, while candidate shocked and isobaric regions and mixes of these features are evident beyond the cavities to the north and south.
    }
    \label{fig:phoenix}
\end{figure*}

X-arithmetic (Fig.~\ref{fig:phoenix}) reveals several candidate-isobaric structures within Phoenix, which could be the result of gas sloshing. They seem to follow a spiral morphology, as found in previous works, and are surmised to be due to a past minor merger event \citep{McDonald_2015, Raja_2020}. There also appears to be a shock front to the north of the AGN/starburst, which is shaped a bit like a handlebar mustache in the ``no isobaric'' panel. Patches of candidate-shocked gas are located in the north, along with a potential bubble at the same location as the proposed ghost cavity. 
However, these features are challenging to identify due to low statistics and the overall surface brightness asymmetry, which complicates our ability to confidently determine whether the model is introducing bias in the interpretation of several features (i.e., the isobaric features to the south).
To be as conservative as possible, we use question marks to denote features which depend on minor adjustments to the global model.

\subsection{Cluster: MS 0735.6+7421}
MS 0735.6+7421 is a galaxy cluster famous for hosting a powerful AGN outburst from what could be one of the largest black holes in the visible Universe \citep{McNamara_2009}. This cool-core cluster is relatively hot (3.5-5\,keV within 100\,kpc) and hosts two enormous cavities, also called ``fat bubbles'' \citep{Sternberg_2009}, coincident with radio emission \citep{McNamara_2005,Vantyghem_2014, Biava_2021, Begin_2023}. Surrounding the cavities is a weak shock front {\citep{McNamara_2005, Vantyghem_2014, OrlowskiSherer_2022}}.

With X-arithmetic (Fig.~\ref{fig:phoenix}), the upper cavity of the cluster can be confirmed as a bubble at all extraction temperatures and abundances. The lower cavity can be identified as a bubble when an extraction temperature around 5.5\,keV is used, but not at lower temperatures. As this higher temperature is better suited to that radius, we identify this feature confidently. The central structure, which becomes clearer after the removal of the global model, looks isobaric at lower extraction temperatures and higher abundances best suited to the region. Portions of the outer elliptical structure, the X-ray enhancement surrounding the cavities beyond 200 kpc radius, can be tentatively confirmed as shocked gas, although their appearance depends heavily on extraction temperature. The rest of the ellipse appears to be a mix of features.

\subsection{Cluster: A2597}
A2597 is a cool-core cluster with evidence of cavities, gas entrainment, and star formation \citep{Tremblay_2012, Tremblay_2018}. Within 20\,kpc of the central AGN, a cold filament ($\sim2.2$\,keV) is present to the northeast, coincident with 1.3-GHz radio emission, and a hot arc ($\sim3.4$\,keV, most prominent in temperature maps) to the west separates the central source and the western cavity \citep{Tremblay_2012}. An entire system of cavities is evident in the inner $25''$, along with tendrils of multiphase gas \citep{Tremblay_2018}. The western cavity is coincident with 330-MHz radio emission \citep{Tremblay_2012}. 

\begin{figure*}
    \centering
    \includegraphics[width=0.85\linewidth]{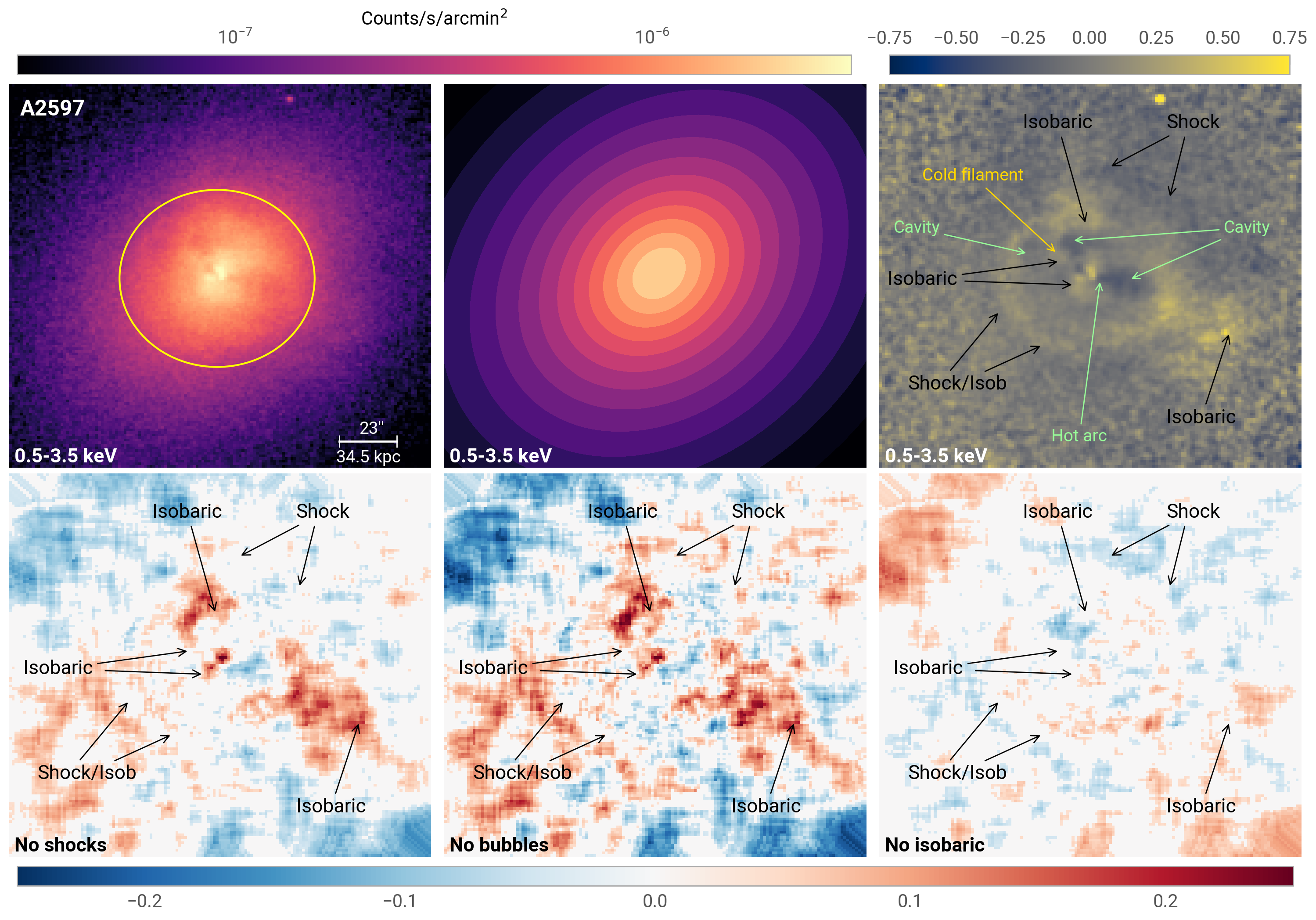}
    \includegraphics[width=0.85\linewidth]{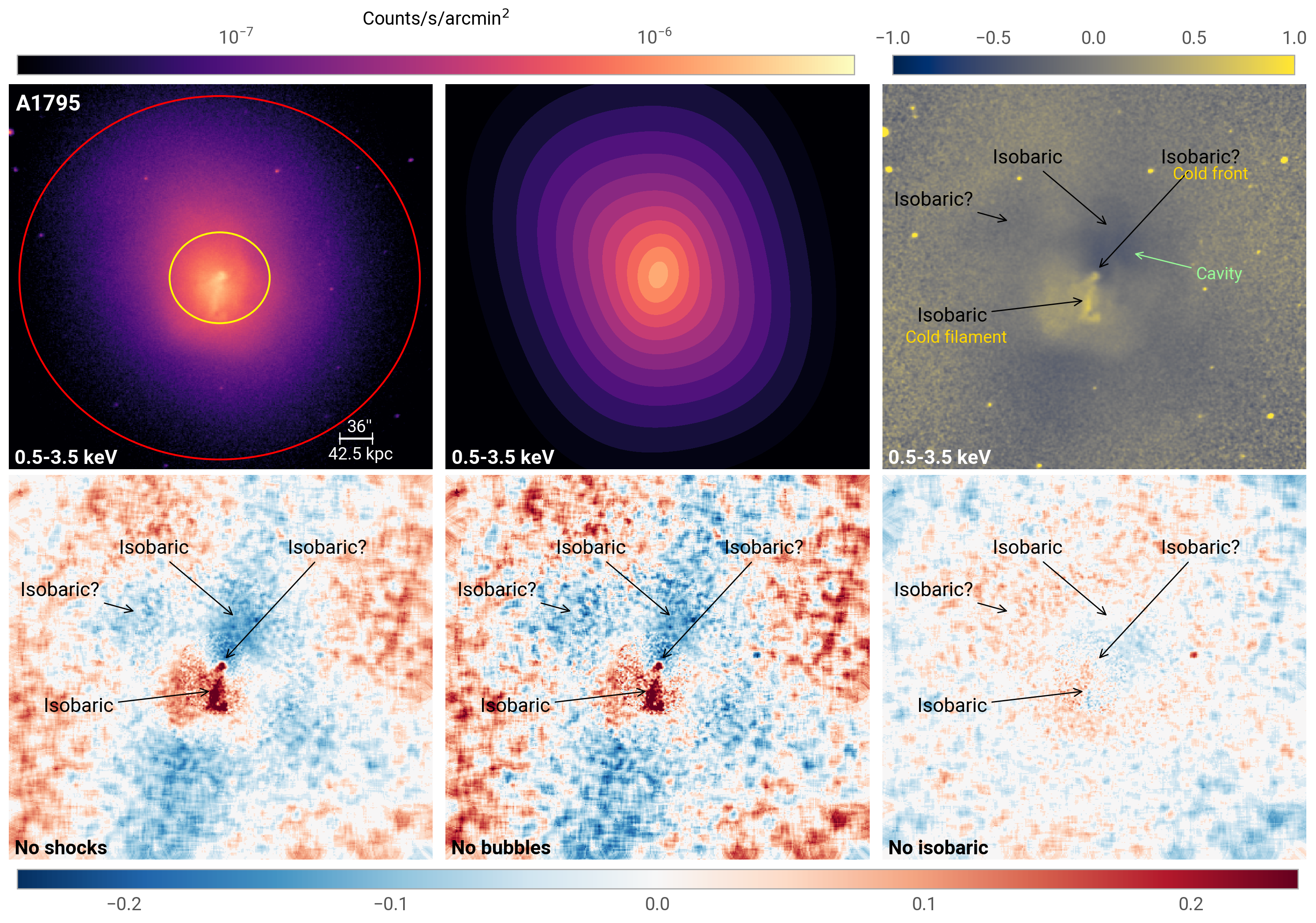}
    \caption{Application of X-arithmetic to A2597 (top) and A1795 (bottom). The panel order and notations are the same as in Fig.~\ref{fig:a2052}.
    In A2597, we find that the cold filament is isobaric, as are other structures to the northwest of the cold filament and to the southwest of cluster center. The ring at around 50\,kpc consists of a shocked feature and mixed shock and isobaric feature to the northwest and southeast respectively.
    In A1795, the cold front and cold filament are both isobaric in nature. Additional isobaric structures are found to the north of these features.
    }
    \label{fig:a2597}
\end{figure*}

With our method (Fig.~\ref{fig:a2597}), we are unable to isolate any of the bubbles identified in other works. However, we do see that the northern cold filament is isobaric. There is an additional isobaric structure to the southwest, beyond the cavities, that has not been studied before. The residual images of the cluster show a ring of emission around the central region and cavities. With X-arithmetic, the northwestern portion of that ring appears to be a shock front, while the southeastern side is a mix of shocks and isobaric structures.

\subsection{Cluster: A1795}
The center of A1795 houses a number of cold filaments, spatially coincident with H$\alpha$ emission, and a surface brightness discontinuity that is potentially a cold front arising from the northern motion of its BCG \citep{Markevitch_2001, Ehlert_2014}. A single cavity is present further to the outskirts in the northwest, but is not associated with any radio emission \citep{Birzan_2020}. For this reason, it is considered a potential ghost cavity, whose counterpart (on the opposite side of the core) may not be visible due to projection effects \citep{Walker_2014}.

Although images are largely featureless (Fig.~\ref{fig:a2597}), making for difficult identification, X-arithmetic is able to isolate the central cold filament as an isobaric structure. A dark region in the residual image to the north, the one that is allegedly due to a galaxy outburst, also appears isobaric. Low statistics in the outer region allow for tentative identification of another isobaric region, possibly related to sloshing.

\subsection{Cluster: A3847}
A3847 is a cluster whose central galaxy has FR II radio morphology. A huge pair of X-ray cavities were detected in the north-south direction, which are not coincident with radio emission \citep{Vagshette_2017}. Potential shock fronts were identified to the north, east, and west, with another weak shock possible to the south. The tentative east and west shocks are closer ($50-55''$) to the center than the north and south shocks ($60-70''$) \citep{Vagshette_2017}.

\begin{figure*}
    \centering
    \includegraphics[width=0.85\linewidth]{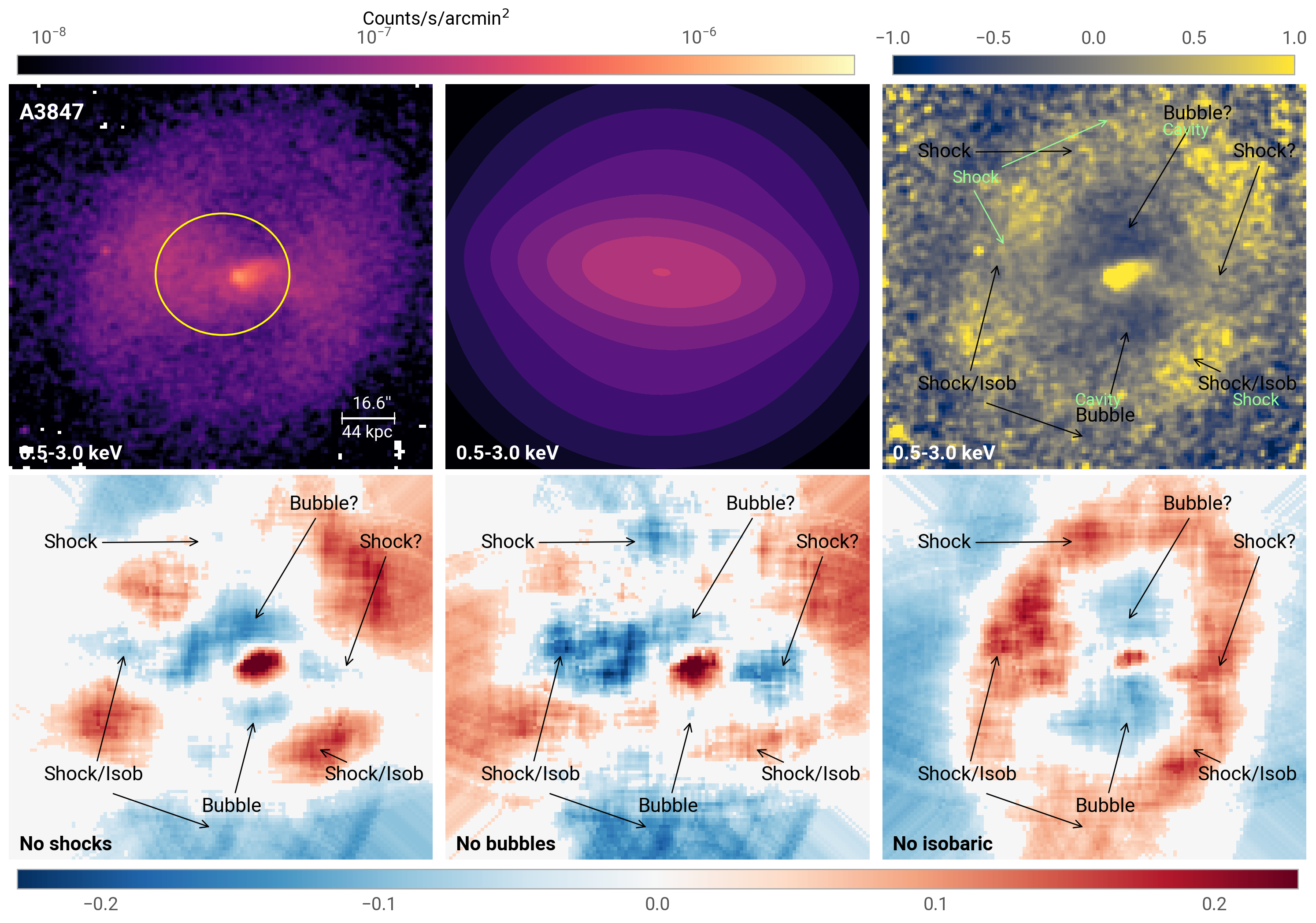}
    \includegraphics[width=0.85\linewidth]{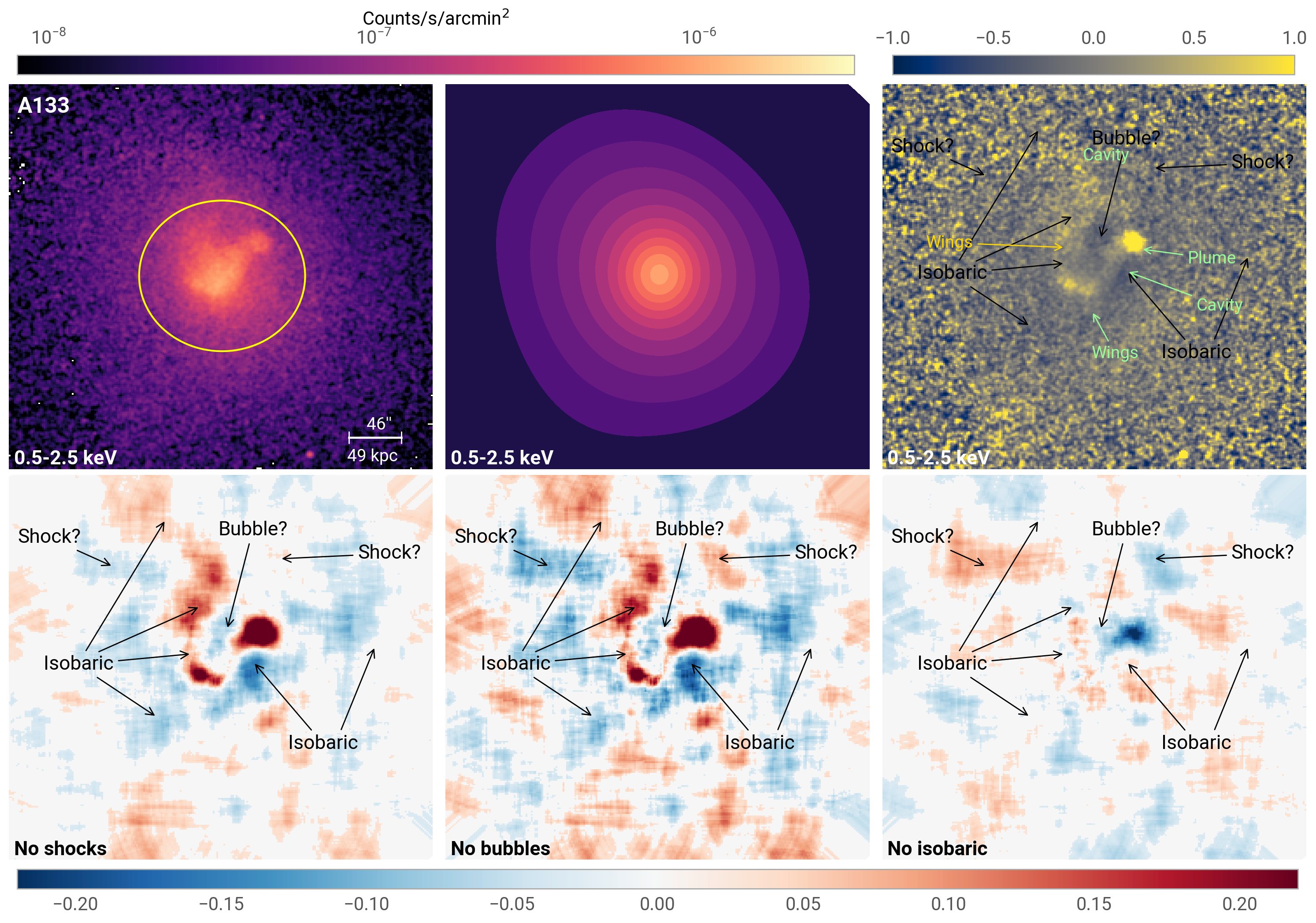}
    \caption{Application of X-arithmetic to A3847 (top) and A133 (bottom). The panel order and notations are the same as in Fig.~\ref{fig:a2052}.
    In A3847, we can confirm the southern bubble and shock to the northeast. We also find that portions of the elliptical feature previously identified as a shock at radius 100\,kpc are a mix of shocked and isobaric perturbations. 
    The so-called wing of A133 is an isobaric feature, as are perturbations to the west of the wing. 
    }
    \label{fig:a3847}
\end{figure*}

Spectral analysis revealed best-fit abundances ranging from $0.1-1.0 Z_{\odot}$ in the central 100 arcseconds \citep{Vagshette_2017}; choice of metallicity has a strong effect on the identification of features with X-arithmetic (Fig.~\ref{fig:a3847}). The southern bubble can be confirmed regardless of the choice of best-fit model. The northern bubble is dependent on model choice and temperature. The surrounding emission appears to generally be a mix of isobaric and shocked perturbations, much like the findings for A2597 and MS 0735. 

\subsection{Cluster: A133}
A133 is considered a disturbed cool-core cluster \citep{Fujita_2002}. Its central emission is often described as having a ``bird-like'' morphology, with a plume extending from symmetric wing features, and radio emission capping the plume \citep{Randall_2010}. On either side of the plume in the north and west, there are depressions interpreted as cavities which are coincident with radio emission \citep{Randall_2010}.

With our method (Fig.~\ref{fig:a3847}), the northern cavity can be confirmed as a bubble (although the region is rather small) but the western cavity appears as an isobaric feature, perhaps related to gas uplift or sloshing. The northern wing and much of the ``tail" of the plume are isobaric, just like the similar looking feature in A1795. An isobaric depression to the east could also be related to sloshing. Candidate shocked gas appears north of the other features, at the location of an edge identified in \citet{Randall_2010}.

\subsection{Cluster: Perseus}
\begin{figure*}
    \centering
    \includegraphics[width=0.85\linewidth]{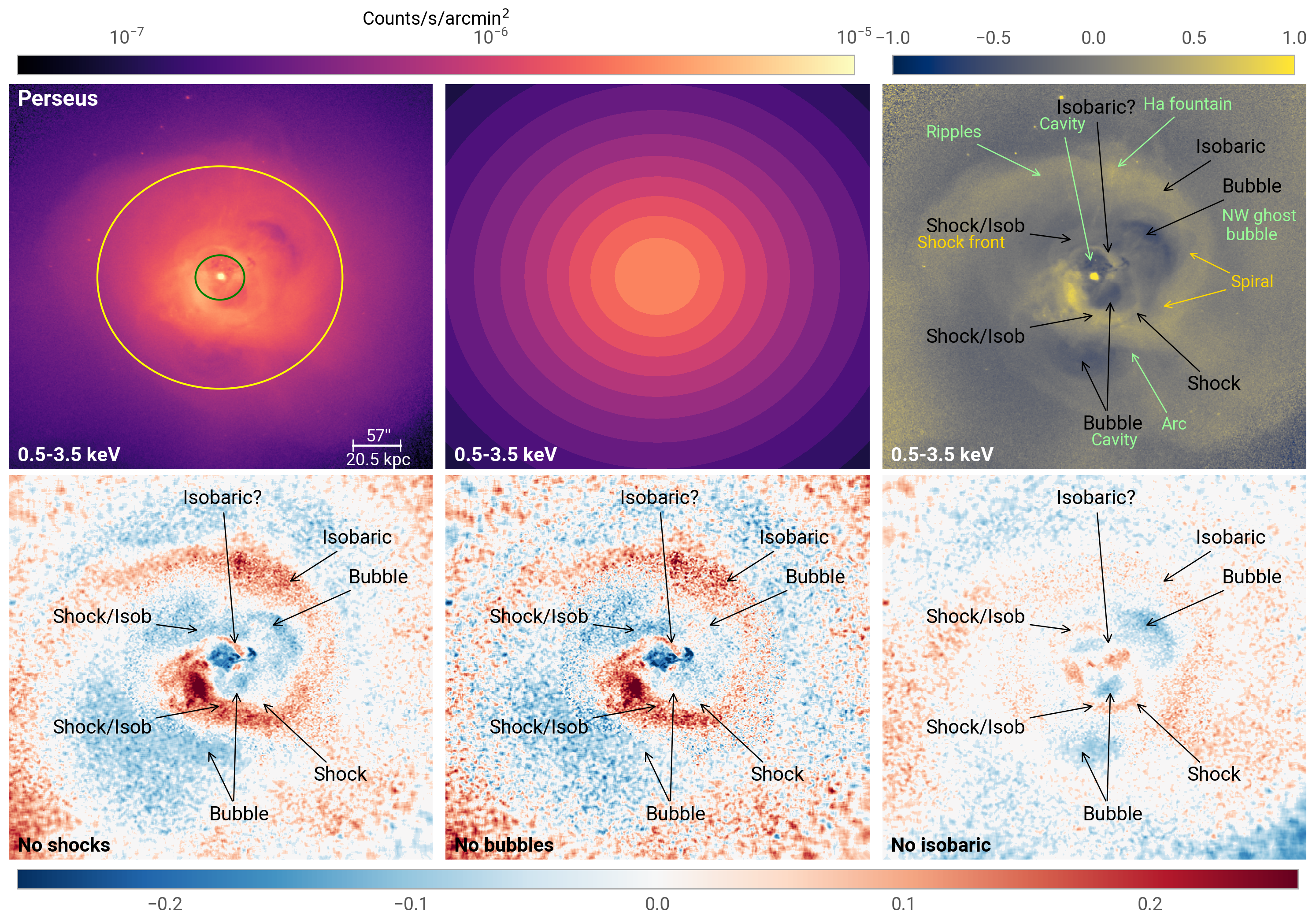}
    \includegraphics[width=0.85\linewidth]{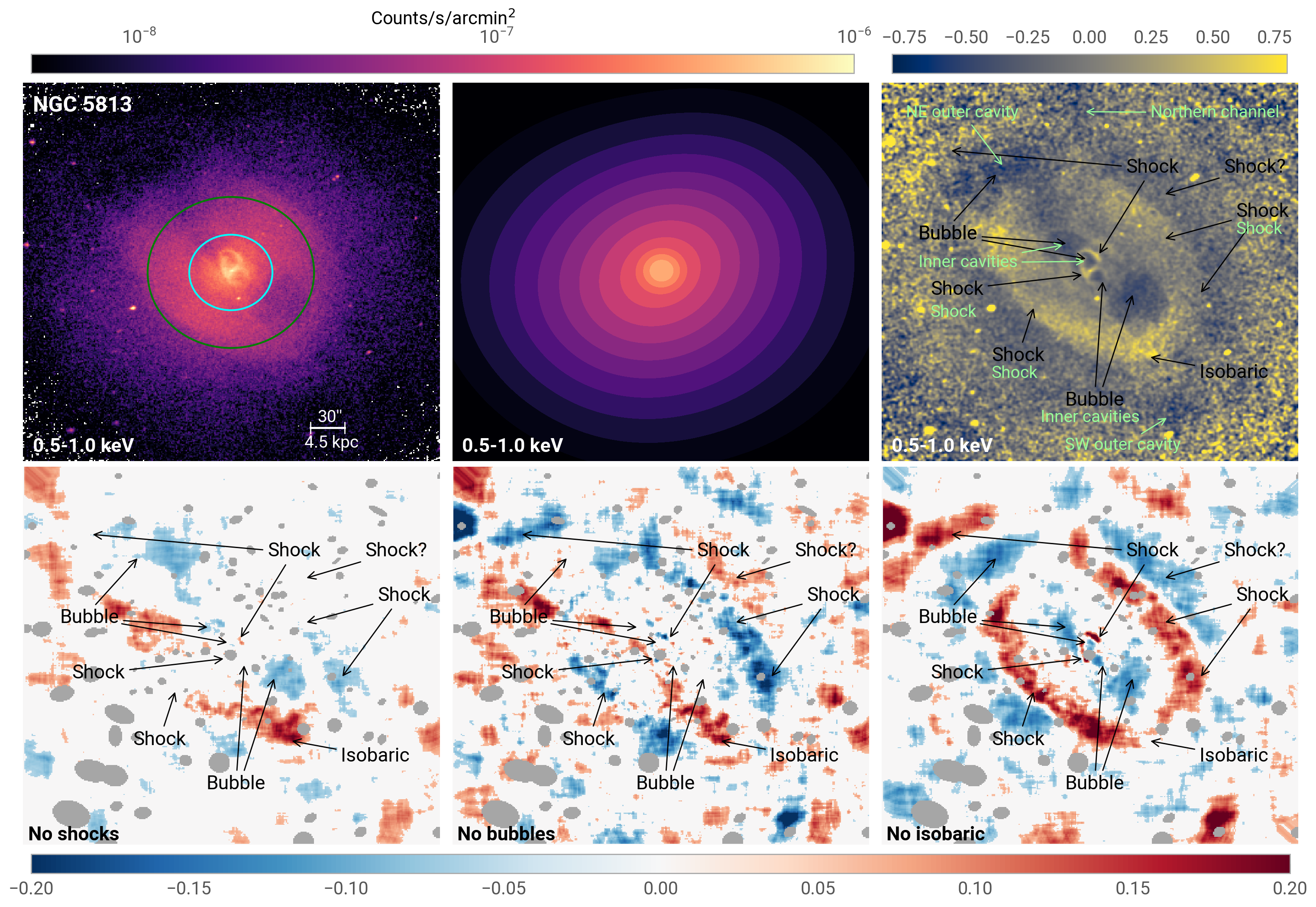}
    \caption{Application of X-arithmetic to Perseus (top) and NGC 5813 (bottom). The panel order and notations are the same as in Fig.~\ref{fig:a2052}.
    In Perseus, we confirm one cavity to the south of the AGN and a ghost bubble each to the southeast and the northwest. We find a clear shocked perturbation to the south of the southern cavity and additional isobaric and shock/isobaric perturbations within 50\,kpc radius around the AGN.
    In NGC 5813 we confirm multiple generations of cavities and shock perturbations. We find an isobaric feature at the radius of the second generation of shocks ($\sim 10$\,kpc).
    }
    \label{fig:perseus}
\end{figure*}

The Perseus Cluster is extremely well-studied \citep[e.g.][]{Fabian_2003, Fabian_2006, Sanders_2007, Fabian_2011}, and has been examined with the X-arithmetic method once before \citep{churazov_2016}. {As the brightest cluster in the X-ray sky, Perseus provides the opportunity to study a large variety of structures that may be less discernible in less luminous or higher redshift objects.} Perseus has two central cavities around its AGN, as well as northwest and southeast ghost bubbles from a prior outburst. There is a prominent spiral structure from gas sloshing and a shock front to the northeast. An H$\alpha$ fountain spouts to the north in the vicinity of a series of ripples, which can additionally be found in the south, east, and west. Only a portion of these ripples is labeled in Fig.~\ref{fig:perseus} due to limited space; \citet{Sanders_2007} show the locations of these features in full depth.

In our revised analysis (Fig.~\ref{fig:perseus}), we find many of the same structures as \citet{churazov_2016}. Just as in that analysis, we find three bubbles: one close to the AGN and two ghost bubbles. The southernmost bubble requires a higher extraction temperature (one that is more consistent with the temperature in that region) than the default to be visible. We also identify a shock south of the inner cavity, and an isobaric spiral structure. The area between the southern cavity and the shock front appears to be a mix between shocked and isobaric fluctuations. The northern shock feature also appears as a mix in this work. The strong (isobaric) spiral arm feature complicates the analysis of the Perseus cluster, leading to many features that appear to be a mix.

\subsection{Group: NGC 5813}

NGC 5813 is a fairly isolated galaxy group whose ICM shows no recent signs of merger activity. \citet{Randall_2011} and \citet{Randall_2015} previously studied the Chandra observations of the group, finding evidence for three distinct AGN outbursts, each with a pair of cavities. Each set of cavities is surrounded by an elliptical shock front with a clear temperature jump. To the northeast, a channel of decreased surface brightness connects to the outermost cavity, a feature which the authors suggest may be energy leaking from the cavity to heat the ICM.

In the X-arithmetic images (Fig.~\ref{fig:perseus}), we are also able to see three generations of cavities. Depending on the choice of extraction temperature, either the outermost cavities or the innermost cavities become more defined, while the middle outburst is stable against choice of temperature. We also see evidence for three corresponding shock fronts for each outburst. The determination of the innermost shock front is sensitive to temperature, while fragments of the other two shock fronts are stable to the choice of extraction temperature and abundance. A feature to the NW outside the second generation shock front is also candidate-shocked gas, but could simply be the result of modeling the steep jump at the shock front as continuous, resulting in spurious identification directly beyond exceptionally bright shocks. There are not enough photons to determine the nature of the regions described as the northern channel and the southwest outer cavity in \citet{Randall_2015}. A surface brightness enhancement in the residual image which extends from the middle shock front to the south is characterized as an isobaric perturbation with this method. It is stable against all systematics.

\subsection{Group: NGC 5044}
NGC 5044 is an asymmetrical X-ray bright group which hosts many small cavities, a semi-circular cold front to the NW of the core, an inner cold front to the SE, two spiral arms, and cool filaments \citep{Buote_2003, David_2009, Gastaldello_2009, OSullivan_2014}. Using MeerKAT data, \citet{Rajpurohit_2025} find diffuse radio emission overlapping, but not confined to, several of the X-ray cavities. The structures present in NGC 5044 are likely to have formed from the combination of AGN feedback and the sloshing motion of the BCG relative to the group gas.

\begin{figure*}
    \centering
    \includegraphics[width=0.85\linewidth]{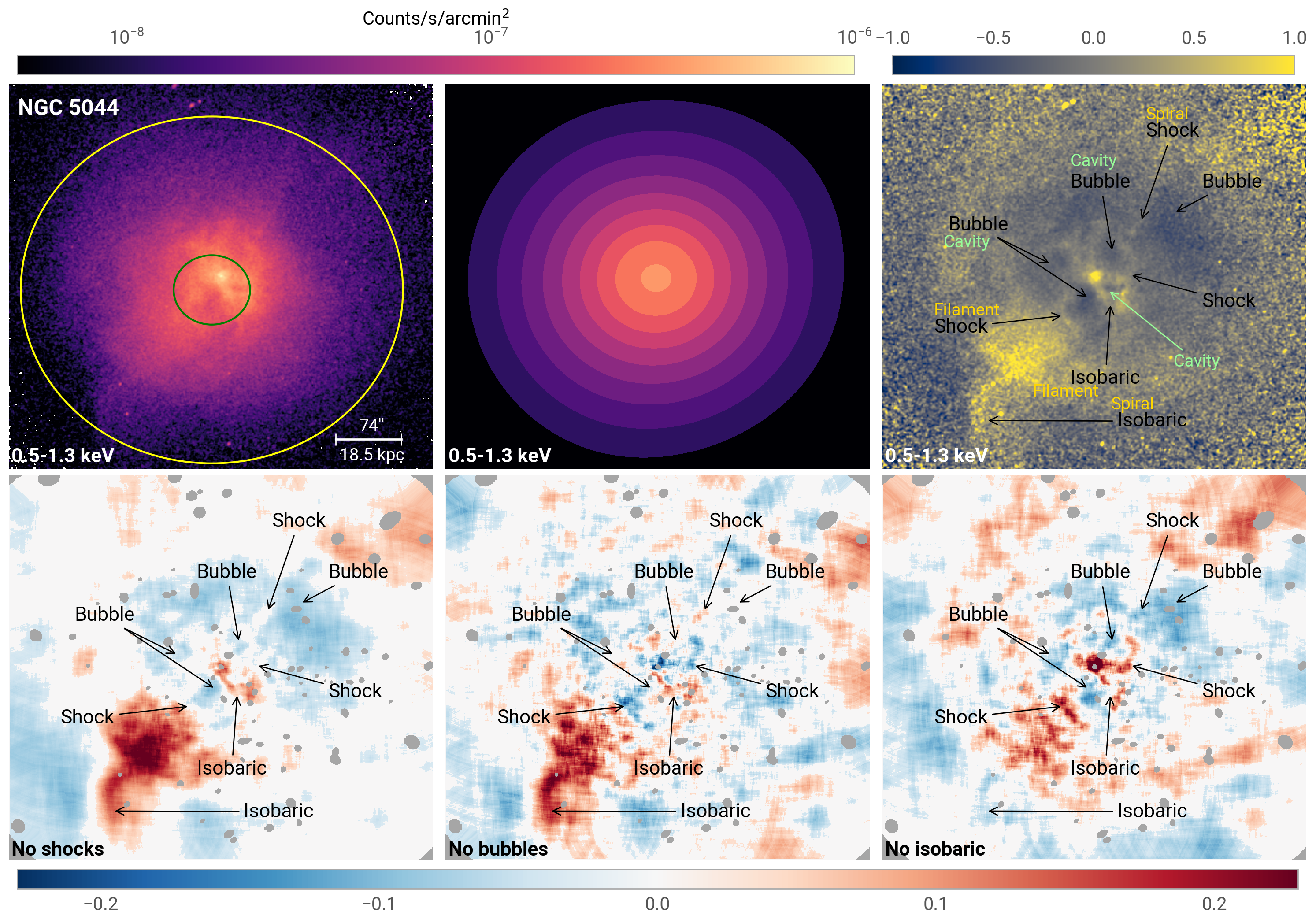}
    \includegraphics[width=0.85\linewidth]{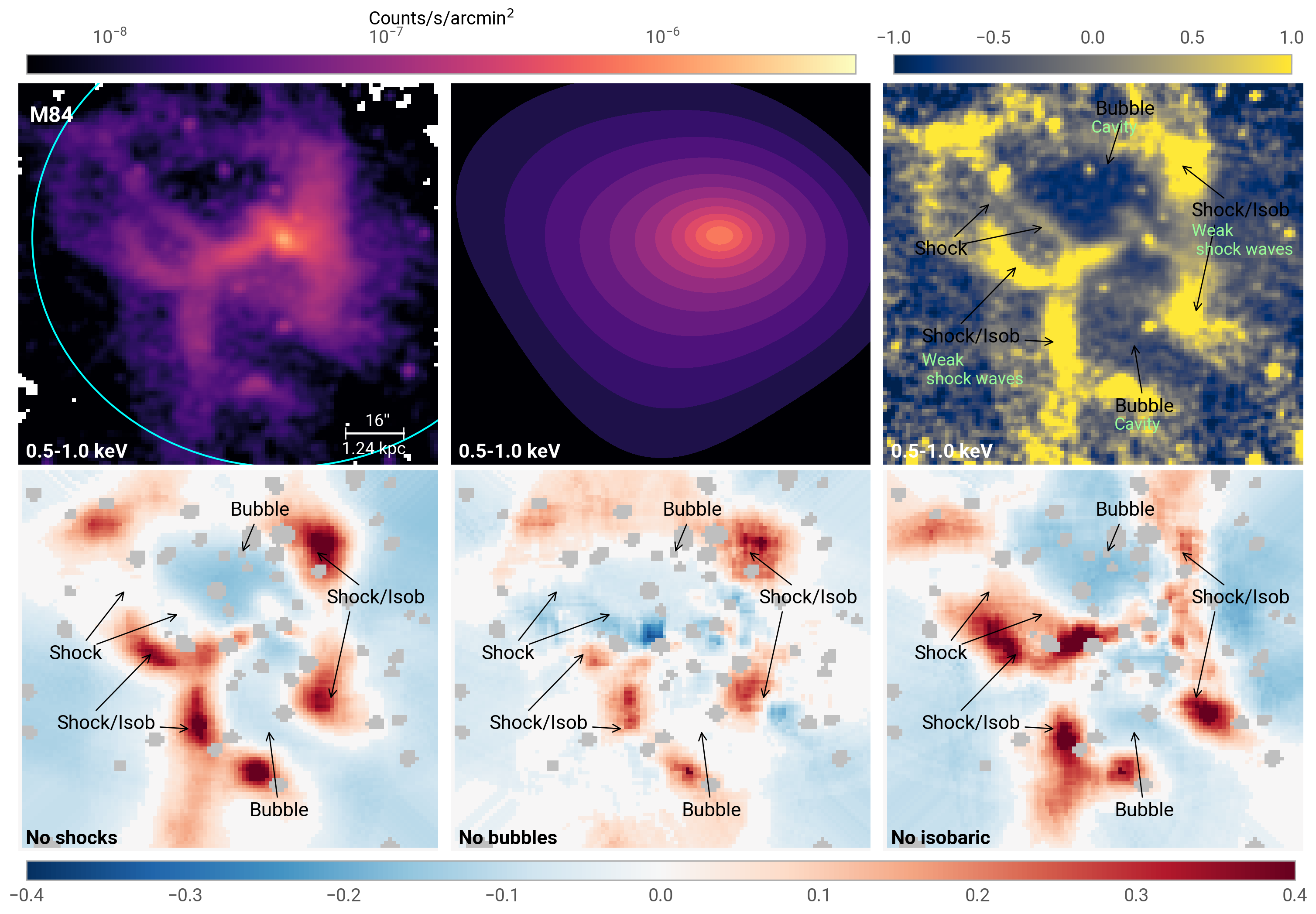}
    \caption{Application of X-arithmetic to NGC 5044 (top) and M84 (bottom). The panel order and notations are the same as in Fig.~\ref{fig:a2052}.
    In NGC 5044, we confirm three bubbles and discover a new one to the northwest. A portion of one so-called filament has a shock signature, while another appears isobaric. The spiral feature to the south appears isobaric, but the spiral to the north appears shocked.
    M84's ``weak shock waves'' are a mix of shocked and isobaric perturbations. A northern and southern cavity are confirmed as bubbles.
    }
    \label{fig:ngc5044}
\end{figure*}

The many point sources in the FOV complicate identification of structures within NGC 5044 (Fig.~\ref{fig:ngc5044}), but some structures are still discernible. Many regions appear to be a mix of all perturbations. The southern spiral arm identified in previous works appears isobaric. The ``H'' filament structure around the central cavities appears to be a combination of shocks and isobaric, reminiscent of the central filaments in NGC 5813. Several depressions have a bubble signature.

\subsection{Galaxy: M84}
M84 (NGC 4374) is a massive elliptical galaxy that features an H-shaped structure around two of its four X-ray cavities \citep{Finoguenov_2008}. It has an FR I class radio jet and is a satellite member of the Virgo Cluster \citep{Bambic_2023}.

Although it is quite difficult to fit a global model, given its asymmetry, features which are robust against choice of model (labeled in Fig.~\ref{fig:ngc5044}) include a lower bubble, along with shocks along the eastern arm of the ``H". The rest of the filament structure is a mix of isobaric and shock, much like similar structures seen in other less massive systems like NGC 5044 and NGC 5813. The extent of the upper bubble that can be confirmed as a bubble with this method is temperature- and model- dependent, with higher temperatures than the default extraction temperature showing more of the bubble than is evident in Fig.~\ref{fig:ngc5044}.

\section{Discussion\label{sec:discuss}}

\subsection{Spectroscopic follow-up}

Although X-arithmetic can be used as a quick identification method of the physical nature of structures, it can also be a tool to optimize spectroscopic analysis. Typically, this analysis requires experimentation with the region choice to locate edges. This process is made challenging by the mix of structures in areas of feedback. X-arithmetic enables easier identification of the edges and extent of structures, and therefore expedites spectral analysis. 

\begin{figure}
    \centering
    \includegraphics[width=0.95\linewidth]{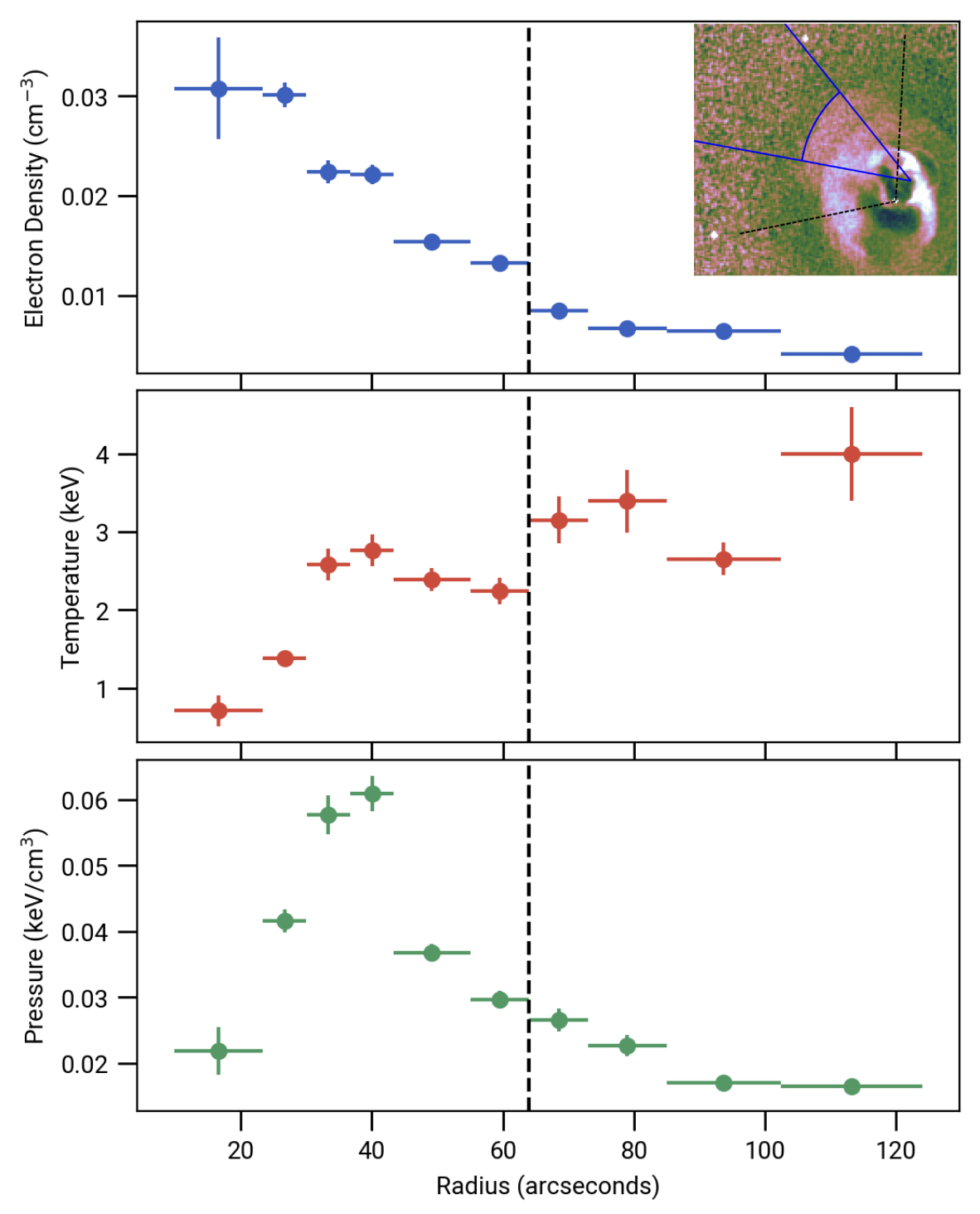}
    \caption{Deprojection analysis of the northeast isobaric feature in A2052. \textit{Inset: }0.5-3.5\,keV band residual image with the spectroscopically analyzed wedge superposed in blue. The best matching sector from \citet{Blanton_2011} is shown as a dashed black wedge.  \textit{Panels: }Deprojected number density of electrons, temperature, and pressure profiles. Radius is measured from the center of the extracted wedge. The position of the edge is marked by a vertical line. There are jumps in temperature and density, but pressure remains continuous across the isobaric feature.}
    \label{fig:sec_shock}
\end{figure}

As an example of X-arithmetic assisting in spectroscopic investigation, and as a demonstration of the method's success, we analyzed, spectroscopically, several features in the Abell 2052 cluster identified through our X-arithmetic process. The first of these is the feature to the northeast of the AGN in A2052 which was identified in \citet{Blanton_2011} as a potential second shock, but appeared isobaric with X-arithmetic. The wedge chosen for the analysis is centered on (RA, DEC)=(229.1828, 7.0245), spans $130-170^{\circ}$, and is pictured in Fig.\,\ref{fig:sec_shock}. Of the sectors explored in \citet{Blanton_2011}, it most closely resembles the NE sector. The perturbation maps where adiabatic and isobaric perturbations have been removed were used to identify the edge of the structure and select the wedge. The identification of the region is sensitive to the exact choice of edge; this choice is simplified by the use of the X-arithmetic maps to identify exactly where the feature ends in the ``no shocks'' image. The wedge was divided into a series of concentric annuli from which spectra were extracted. Each spectrum was fit simultaneously with an absorbed APEC model using \texttt{projct} from XSPEC \citep{Arnaud_1996}, which performs a 3D to 2D deprojection. This resulted in the deprojected temperature, density, and pressure profiles shown in Fig.\,\ref{fig:sec_shock}. 
Across the identified edge, the temperature and density show jumps, while the pressure is continuous, supporting the isobaric nature of the feature revealed with X-arithmetic.
The colder temperature inside the edge is indicative of a cold front, perhaps related to gas sloshing. At smaller radii (around 40'' in Fig.~\ref{fig:sec_shock}), the inner shock identified by \citet{Blanton_2011} is also visible in this wedge as a clear pressure jump. This is also consistent with the X-arithmetic identification.

\begin{figure}
    \centering
    \includegraphics[width=0.95\linewidth]{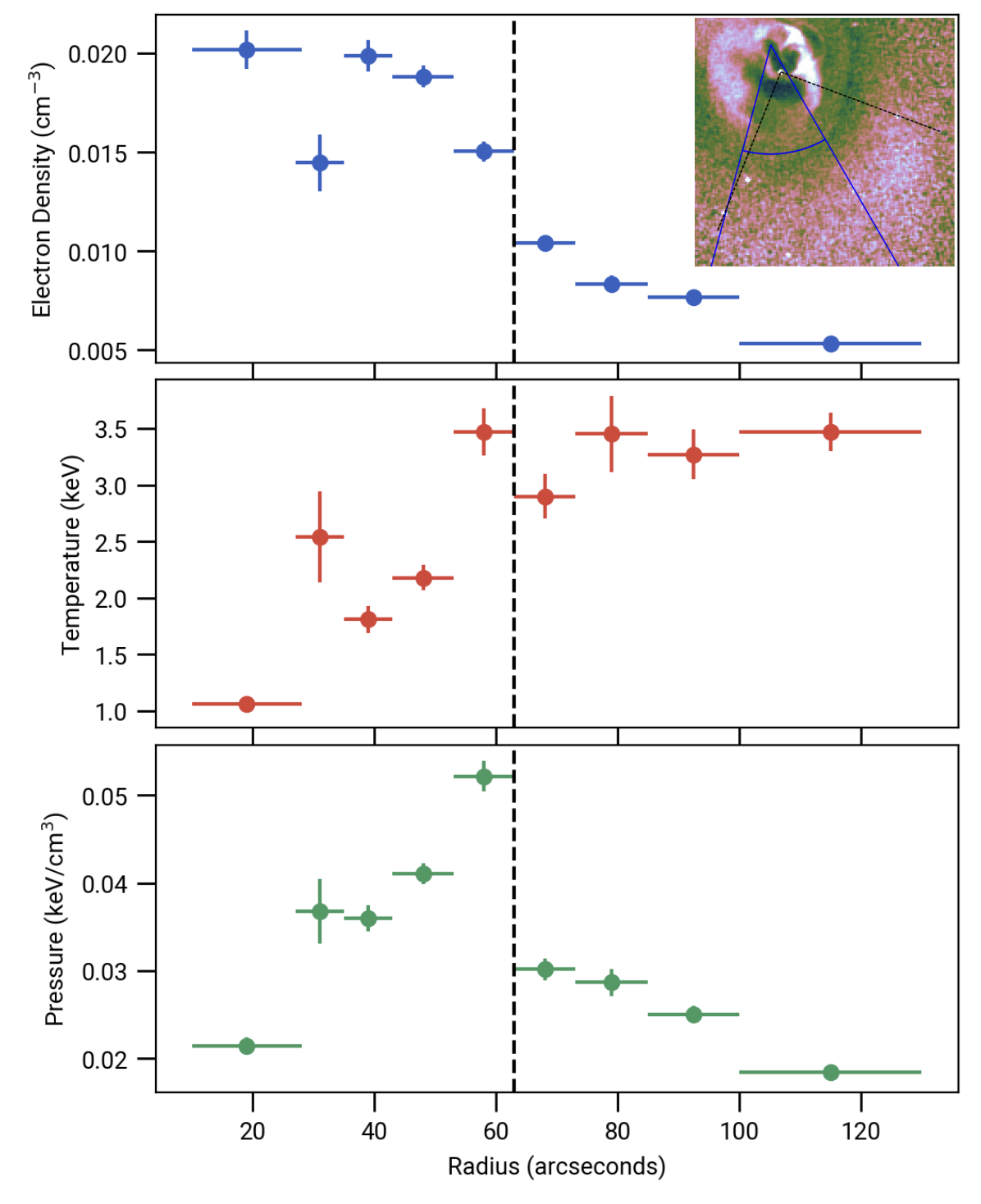}
    \caption{Deprojection analysis of the shock feature in A2052. The panel order and inset are the same as in Fig.~\ref{fig:sec_shock}. 
    Temperature, density, and pressure drop across the shock front. }
    \label{fig:first_shock}
\end{figure}

The second A2052 feature examined spectroscopically is a segment of the so-called ``first shock'' that is prominent in X-arithmetic images. This segment was extracted south of the AGN, where there is less mixing of shocked and isobaric structures than in other regions. The shock edge was identified using the maps with removed adiabatic and isobaric perturbations. These were used, as before, to select the region for analysis, which is centered at (RA, DEC)=(229.1869, 7.0259), spans $255-300^{\circ}$, and is displayed in Fig.\,\ref{fig:first_shock}. This region most closely matches the SW sector in \citet{Blanton_2011}. The deprojected profiles of Fig.\,\ref{fig:first_shock} show jumps in all characteristics; the temperature and pressure rise inside the edge and drop across it, while the density drops. This behavior is characteristic of shock fronts. 

\begin{figure}
    \centering
    \includegraphics[width=0.95\linewidth]{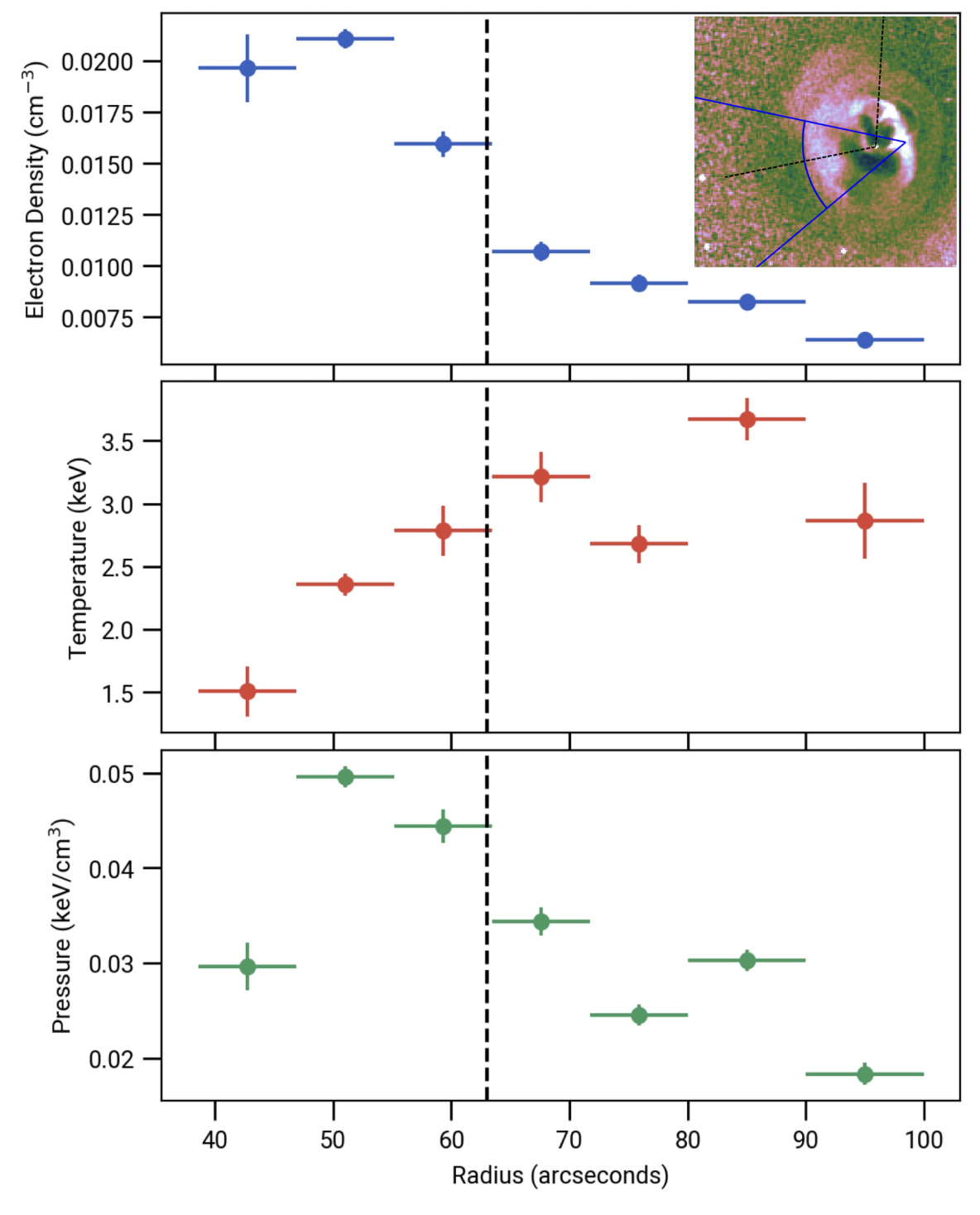}
    \caption{Deprojection analysis of the ``isothermal shock'' feature in A2052. The panel order and inset are the same as in Fig.~\ref{fig:sec_shock}. 
    Density and pressure drop across the shock front, but a temperature jump is not present. This implies a mix of adiabatic and isobaric processes in this region that combine to obscure any temperature jumps.}
    \label{fig:iso_shock}
\end{figure}

The third and final feature in A2052 explored here is the shocked region to the southeast. Spectroscopic analysis of the southeast area was not reported in \citet{Blanton_2011}, but our chosen wedge has some overlap with the NE sector of that work, where the inner shock was identified.
Following the same procedure as with the other two features, the edges were identified using the ``no shocks'' and ``no isobaric'' maps, and the deprojected profiles were extracted from a wedge region that spans $168-220^{\circ}$ and is centered on (RA, DEC)=(229.1804, 7.0222) (see Fig.~\ref{fig:iso_shock}). 
This sector contains a jump in both density and pressure across the edge, but no jump in temperature. This implies a mix of adiabatic and isobaric processes in this region, which could be responsible for masking the temperature across the shock.
This matches the behavior of a feature originally identified in the Perseus Cluster by \citet{Fabian_2006}, referred to as an isothermal shock; other works have found similar features in other clusters \citep{Sanders_2006}.

Although there are many additional features discovered in this work which could be subjected to spectroscopic follow-up, these examples serve to illustrate the power of X-arithmetic and how it can be used to guide further analysis.

\subsection{Application to simulations}\label{subsec:sims}

{Beyond observations, we applied the X-arithmetic method to a numerical simulation of a massive halo to verify that the method performs as intended and to show that its application to simulated data is straightforward. The latter is particularly important as it opens a future opportunity to efficiently assess how well simulations capture the details of feedback physics. Typically, AGN feedback is modeled in cosmological simulations as kinetic and/or thermal energy injection. The type, duration, and amount of energy injected differs between simulations. These approaches have proven to suppress star formation and are often tuned to reproduce optical properties of host galaxies \citep{Bower_2006, Croton_2006, Dubois_2013, Somerville_2015, Weinberger_2018, Nelson_2021}, but how well the gas physics is captured is not well studied. 
Some works have compared the global X-ray properties of clusters in simulations to observations \citep[e.g.,][]{Barnes_2017, Vogelsberger_2017, Truong_2020}, but none have yet examined the detailed structures. X-arithmetic has the potential to provide new gas characteristics in simulated clusters that can be compared directly to observations. In this work, we apply X-arithmetic to a single simulated halo to demonstrate its application.}

{We have applied X-arithmetic to a halo from TNG300 \citep{Nelson_2019, Pillepich_2018, Springel_2018, Nelson_2018, Naiman_2018, Marinacci_2018}. The chosen snapshot }was selected from TNG300-1 snapshots of halos with mass M$_{200} > 10^{14}$\,M$_{\odot}/h$ at zero redshift (those massive enough to have transitioned to kinetic mode) which show low central entropy and have evidence of feedback. 
X-ray mock images of the galaxy cluster based on the simulated halo were generated using \texttt{yt} \citep{Turk_2011}, \texttt{pyXSIM} \citep{Zuhone_2016}, and \texttt{SOXS} \citep{Zuhone_2023} software. With \texttt{yt}, we apply a filter to the simulation data to select only gas particles hot enough to emit X-rays. 
\texttt{pyXSIM} allows for the selection of a source model (in this case, APEC) which is used to generate a 3D photon population and project it onto a 2D sky plane. 
Finally, we use \texttt{SOXS} to simulate an X-ray instrument response and produce counts images, exposure maps, and fluxed images. 
For this example, we assumed a 500\,ks exposure time with the \textit{AXIS} satellite \citep{Reynolds_2023} response files, a redshift $z=0.0179$, and 0.5-2.5 keV and 2.5-7.5 keV energy bands. X-arithmetic was then applied to the images following the method applied to the Chandra sample as described in Section~\ref{sec:methods}, using a patched single elliptical model with patching radius 100''.

\begin{figure*}
    \centering
    \includegraphics[width=0.7\linewidth]{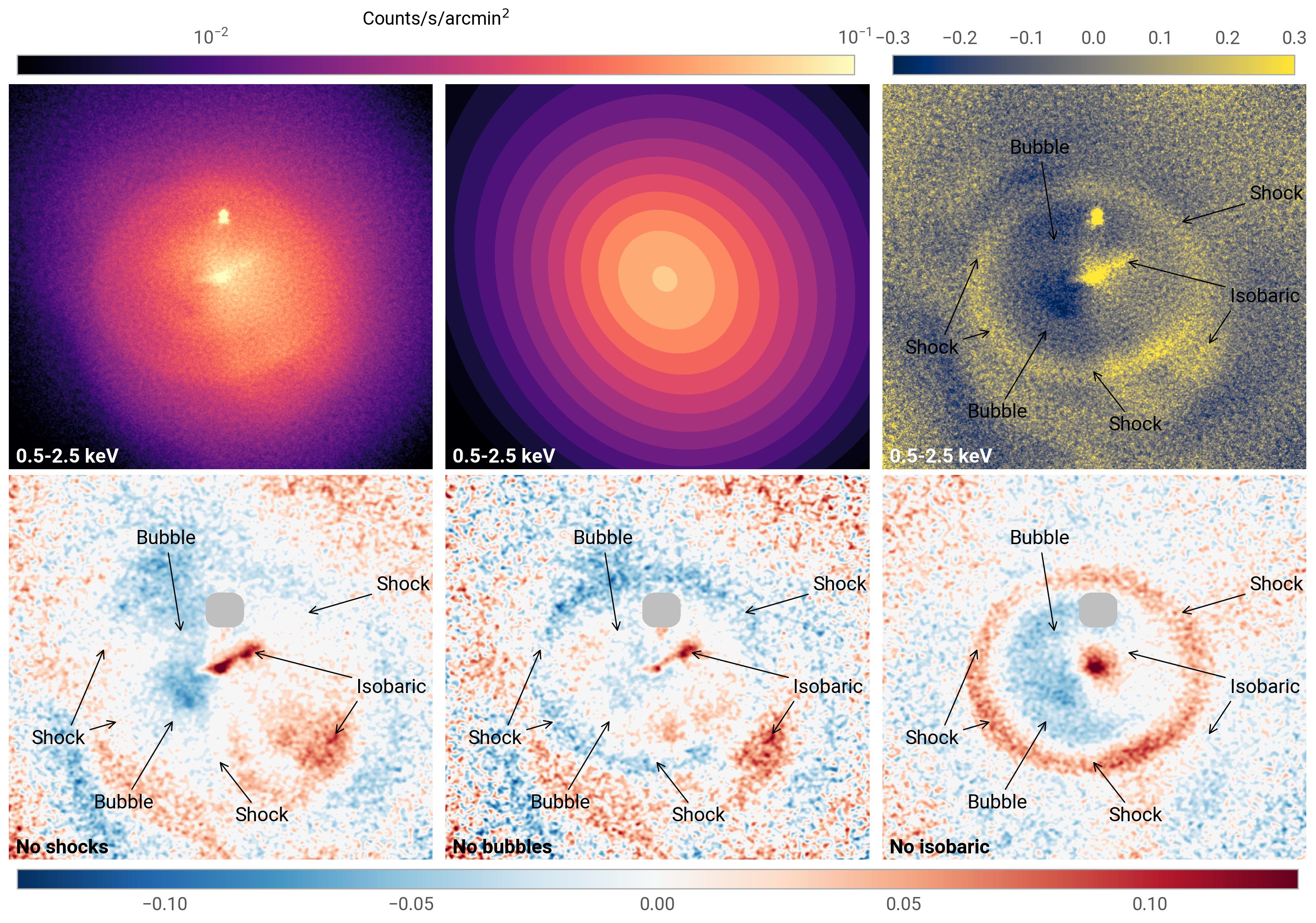}
    \includegraphics[width=0.261\linewidth]{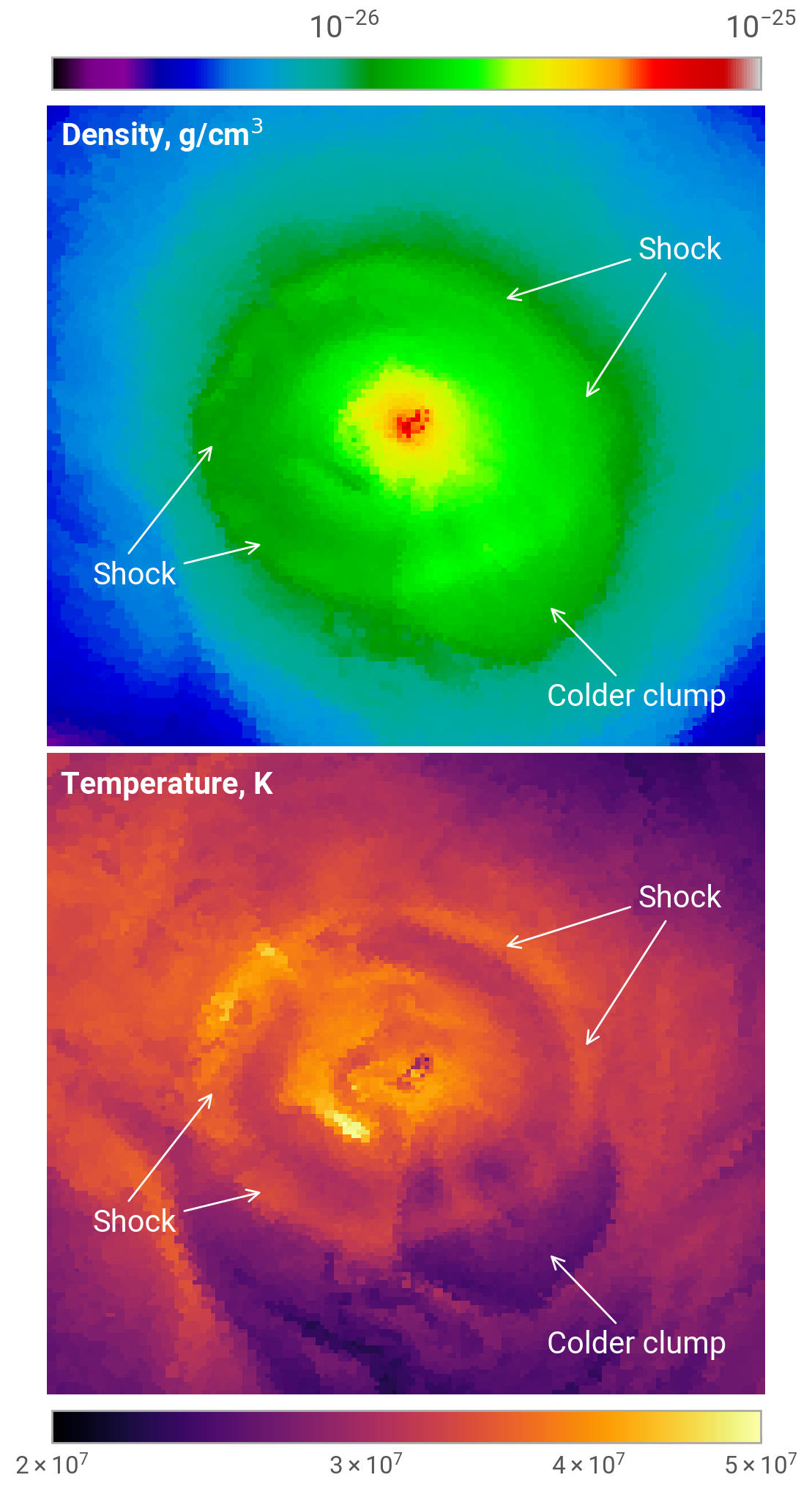}
    \caption{\textit{Leftmost 3 columns:} Application of X-arithmetic to a mock 500\,ks AXIS image of a $3.7 \times 10^{14}$\,M$_{\odot}/h$ TNG300 cluster positioned at z=0.0179. Panel arrangement is the same as in Fig.~\ref{fig:a2052}. With X-arithmetic, we can identify a ring-like shock structure that is disrupted to the SW, and inside the ring, several cavities. \textit{Rightmost column:} Density (top) and temperature (bottom) slices of the simulated cluster, passing through the cluster center, perpendicular to the line of sight. The shock is clear as a density jump in the top slice, and the SW disruption of its identification with X-arithmetic corresponds to a colder clump in its temperature slice.
    }
    \label{fig:tng_sim}
\end{figure*}

The first three columns of Fig.~\ref{fig:tng_sim} show the application of X-arithmetic to the example TNG300 halo. 
A nearly perfect ring-like structure, prominent in the residual image, is clearly shock-like around the majority of its radius. 
An exception is in the SW, where a prominent isobaric feature masks the shock. 
The final column of Fig.~\ref{fig:tng_sim} shows the gas density and temperature slices of the cluster, and offers a clue as to why this might be. 
The shock front, evident as a density and temperature jump, is not spherical to the SW. 
One possible interpretation is that, since the SW region has a colder clump, the shock wave speed is reduced as it traverses the colder medium. 
The comparison of the X-arithmetic findings with the density and temperature profiles offers evidence of the reliability of the method in identifying shocks and isobaric features.
Interior to the shock, two portions of the eastern emission appear to be bubbles. An extension near the center of the halo appears isobaric, and stretches in the opposite direction from the visible cavity, as in A1795.

This example demonstrates that X-arithmetic can be used as an efficient tool for verifying the implementation of feedback physics in simulations. For direct comparison with observations, the analysis of a larger sample of simulated clusters is required, which is straightforward, but beyond the scope of this work.

\subsection{Application to future missions: AXIS}

AXIS is a proposed X-ray imaging observatory in the Probe Explorers mission class that was selected by NASA for additional review \citep{Reynolds_2023}. It will provide arcsecond-resolution imaging in the 0.3-10\,keV band, with 15 times Chandra's effective area at 0.5\,keV, and six times at 6\,keV \citep{Reynolds_2023}. Its combination of large effective area and near constant 1.6 arcsecond angular resolution within the entire field of view makes it the ideal instrument to collect data for X-arithmetic application. These qualities will make it possible to expand future samples to a greater number of gaseous halos whose inclusion in our sample is limited by the available number of photons.

\begin{figure}
    \centering
    \includegraphics[width=0.85\linewidth]{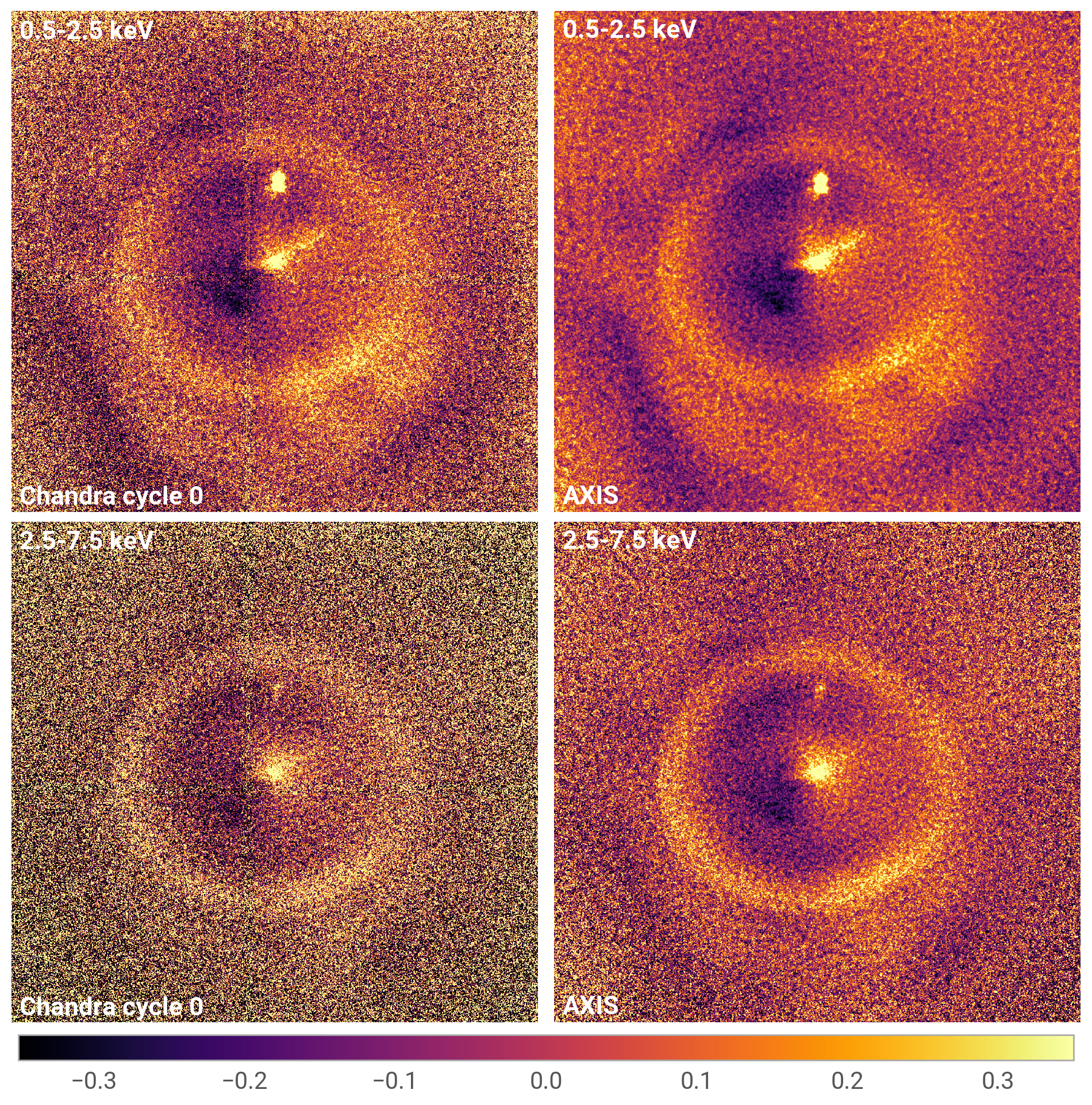}
    \caption{Soft (0.5-2.5\,keV, top) and hard (2.5-7.5\,keV, bottom) band residual images of mock 500\,ks Chandra ACIS-I cycle 0 (left) and AXIS (right) observations of a $3.7 \times 10^{14}$\,M$_{\odot}/h$ TNG300 cluster positioned at z=0.0179. The images are lightly smoothed for visual purposes and displayed with the same scale.}
    \label{fig:chandra_vs_axis}
\end{figure}

In addition to showing an application to simulation data, Fig.~\ref{fig:tng_sim} demonstrates the X-arithmetic feasibility study with AXIS. With sufficient exposure time, the nature of the most prominent structures can be reliably recovered. For a direct comparison with Chandra, Fig.~\ref{fig:chandra_vs_axis} shows mock soft and hard band residual images of the same TNG300 halo presented in Fig.~\ref{fig:tng_sim}. The cycle 0 response files were used to simulate Chandra. The comparison demonstrates the advantage that AXIS will have over Chandra in collecting a greater number of photons across the entire field of view within the same exposure time. Given its stable PSF across the field of view, AXIS images will be ideal for X-arithmetic application, allowing us to probe beyond the brightest central regions.

\subsection{Limitations and trends}

The X-arithmetic method is best suited to bright, prominent structures in gaseous halos. Faint features can be misidentified in halos with irregular morphology due to the dependence on the choice of the underlying model, so caution should be taken in these cases. In this work, we only identified bright features that were stable against the choice of underlying model.
The method works best for identifying weak shocks (adiabatic) and subsonic motions/cooling (isobaric) structures. Bubbles (appear as isothermal) are challenging to identify unless they are large and prominent{, but their presence can be confirmed independently in most cases with the use of radio observations}.

Assessing the overall trend in the sample, we see shocked gas and bubbles more consistently in lower mass objects (galaxies and groups) aka NGC 5813, M84, and NGC 5044. All three objects have multiple shocks within 20 kpc and a similar ``H'' structure around central cavities. This structure is a shocked gas and isobaric mix in NGC 5044 and M84, and exclusively shocked gas in NGC 5813. Massive FR I clusters, in contrast, feature bright isobaric structures around their bubbles and fewer shocks per halo. M87 is an in-between source and has multiple shocks (characteristic of groups) and strong isobaric features (typical for regular clusters). This trend could be due to a more violent form of AGN feedback in groups than in clusters. Alternatively, outbursts of the same power output could merely have a greater effect on groups, which have weaker gravitational potentials and lower gas temperatures, making outbursts more pronounced and detectable. The difference could also reflect a combination of more vigorous radiative losses in cooler systems (groups) and more readily visible changes in the X-ray spectra in response to shocks with similar Mach numbers as in clusters.
 One should also consider that the physical sizes of regions analyzed in clusters and groups differ significantly. It is possible that we simply lack the resolution to capture the diverse structures within the innermost regions of the most massive systems.

\section{Conclusions\label{sec:conclude}}
We have applied the X-arithmetic method to a sample of 15 massive gaseous halos with deep Chandra exposures. We have identified structures of an adiabatic, isothermal, or isobaric nature imprinted on their X-ray images. We focused on cool core regions that are affected by AGN feedback activity. Our main findings are summarized below.

\begin{itemize}
    
    \item We robustly identified shocked gas in 11/15 ($\sim 73 \%$), bubbles in 9/15 ($\sim 60 \%$), and isobaric structures in 13/15 ($\sim 87 \%$) of our sample. Of more massive systems (with temperature $\gtrsim 2$ keV, see Table \ref{tab:sample}), 8/12 ($\sim 67\%$) have shocked gas, 6/12 ($\sim 50 \%$) bubbles and 11/12 ($\sim 92 \%$) isobaric structures. Of less massive systems, 3/3 (100\%) have shocked gas and bubbles and 2/3 ($\sim 67 \%$) have isobaric perturbations. 

    \item In regular FR I clusters with clear feedback activity, the inner 10 kpc regions have prominent isobaric features and no shocks, while groups and galaxies show multiple shocks within the innermost regions and only weak isobaric structures often mixed with shocks. This could suggest a less violent form of AGN feedback in clusters than in groups, or reflect a shallower gravitational potential of groups, meaning a stronger effect of feedback on groups for the same AGN power output, or could be an observational bias due to different sizes of resolved regions in clusters vs.\ groups.

    \item In clusters with large bubbles (MS 0735, Hydra A, A3847), shocks reach large radii, up to 200 kpc, and are often mixed with isobaric structures. We did not identify shocks at such large distances in other clusters that have smaller cavities.
    
    \item We identify several structures whose physical mechanisms were previously unknown: in A2052, the isobaric feature indicative of gas sloshing; in Centaurus, an isobaric strip close to the central AGN; in Hydra A, Phoenix, MS 0735, A2597, A3847, and NGC 5044, the shocked and isobaric structures surrounding the cavities; in A1795 and A133, isobaric structures. The odd structures identified as plumes and arms in Centaurus and M87, respectively, are a shocked gas/ isobaric mix at their densest/brightest parts and exclusively isobaric beyond them.

    \item A133 and A1795 do not show typical features of an active phase of AGN feedback, with only isobaric structures definitively present and neither strong bubbles nor strong shocks identified. Alternatively, it could be that the feedback is present but exceptionally ``gentle'' in these systems.

    \item Cygnus A -- the clearest example of an FR II object in our sample -- shows more cohesive, symmetric shocked gas that reaches large radii compared to the massive FR I clusters.

    \item Although capable of identifying the nature of structures independent of spectral analysis, X-arithmetic can aid in optimization of spectral analyses by enabling clear identification of edges and structure locations.
   
    \item X-arithmetic can be applied to numerical simulations of gaseous halos through the generation of mock X-ray images, or by extending the idea to density and temperature fluctuations directly from simulations. {The application to simulations is straightforward, and therefore has the potential to be used in future to verify specific feedback implementations with a larger sample of simulated clusters.}

    \item In the case of high signal-to-noise halos or simulations, the three X-arithmetic maps can be combined into a single three-color image as in Fig.~\ref{fig:3color} where each perturbation is represented by a color, simplifying feature identification.

\end{itemize}

Future X-ray missions with high-resolution imaging capabilities and large effective area, like AXIS, will reveal the full potential of the X-arithmetic method, allowing the studies of the nature of structures in the ICM beyond the brightest core regions and in larger samples of massive galaxies, groups, and clusters.

\section*{acknowledgments}
This research has made use of data obtained from the Chandra Data Archive and the Chandra Source Catalog, and software provided by the Chandra X-ray Center (CXC) in the application package CIAO \citep{2006SPIE.6270E..1VF}. HM and IZ are partially supported by the NASA Astrophysics Data Analysis Program under grant number 80NSSC24K1488. IZ acknowledges partial support from the Alfred P. Sloan Foundation through the Sloan Research Fellowship. IZ performed part of the work at the Kavli Institute for Theoretical Physics (KITP) supported by grant NSF PHY-2309135. WF acknowledges support from the Smithsonian Institution, the Chandra High Resolution Camera Project through NASA contract NAS8-03060, and NASA Grants 80NSSC19K0116 and GO1-22132X. CZ was supported by the GACR grant 21-13491X. The authors thank John ZuHone for his expertise and assistance in generating the mock X-ray images of the TNG300 halo.

\section*{Data Availability}
This article employs a list of Chandra datasets (Table \ref{tab:obsids}), obtained by the Chandra X-ray Observatory, contained in~\dataset[doi: 10.25574/cdc.414]{https://doi.org/10.25574/cdc.414}.
X-arithmetic map data for the sample of 15 galaxy clusters, galaxy groups, and massive galaxies, as well as a mock observation of a TNG 300 halo are available on Zenodo: \dataset[doi:10.5281/zenodo.15708466]{https://doi.org/10.5281/zenodo.15708466}.

\vspace{5mm}
\facilities{CXO}

\clearpage

\appendix
\section{Additional tables}
As discussed in Section~\ref{sec:data}, Chandra archive observations for each object were used to produce the images. Table~\ref{tab:obsids} lists the Observation IDs used for each object.

\begin{table*}
\begin{tabular}{ll}
       \hline
       Object & ObsIDs\\
       \hline
       M84 & 22195, 22196, 5908, 22164, 22174, 20543, 21867, 22128, 22127, 22177 \\ & 22153, 21845, 22166,  22144, 20542, 22126, 22113, 22176, 20541, 20540 \\ & 22142, 22175, 20539, 21852, 6131, 22163, 22143\\
       
       NGC 5813 & 9517, 12952, 13246, 13247, 13253, 13255, 12953, 12951, 5907 \\
       
       NGC 5044 & 9399, 17195, 17196, 17653, 17654, 17666 \\
       
       M87/Virgo & 5826, 5827, 5828, 2707, 3717, 11783, 6186, 7210, 7211, 7212 \\
       
       A2052 & 10477, 10480, 10914, 10915, 10916, 10917, 5807, 10879, 10479, 10478\\
       
       Centaurus & 16223, 4954, 16609, 16534, 5310, 16607, 4955, 16224, 16608, 16225, 16610\\
       
       A133 & 3183, 3710, 9897, 13518 \\
       A3847 & 11506, 15091, 15092, 15643 \\
       Hydra A & 4969, 4970 \\
       A1795  & 18429, 19868, 19879, 20648,
        21835, 22833, 24603, 25670, 26382, 27806, 6163 \\ &
        13107, 14271, 15491, 16466, 17401, 17683, 18430, 19869, 19880, 20649 \\ & 21836, 22834, 24604, 25671, 27020, 3666, 13108, 14272, 15492, 16467, 17402 \\ &
        17684, 18431, 19870, 19881, 20650, 21837, 22835, 24605, 25672, 27021 \\ & 5286,
        10898, 13109, 14273, 16432, 16468, 17403, 17685, 18432, 19871, \\ & 19968, 20651, 21838,
        22836, 24606, 25673, 27022, 5287, 10899, 13110, 
        14274 \\ & 16433, 16469, 17404, 17686, 18433, 19872, 19969, 20652, 22837, 24607 \\ & 25674, 27023, 5288,
        10900, 13111, 14275, 16434, 16470, 17405, 18423 \\ & 18434, 19873,
        20642, 20653, 21840, 22838, 24608, 25675, 27024, 5289,
        10901 \\ & 13112, 15485, 16435, 16471, 17406, 18424, 18435, 
        19874, 20643, 21830 \\ & 21841, 22839, 24609, 25676, 27025, 5290,
        12026, 13113, 15486, 16436, 16472 \\ & 17407, 18425, 18436,
        19875, 20644, 21831, 22829, 22840, 24610, 25677 \\ & 27026, 6159,
        12027, 13416, 15487, 16437, 17397, 17408, 18426, 18437, 19876 \\ &
        20645, 21832, 22830, 24600, 24611, 25678, 27027, 6160,
        12028, 14268, 15488 \\ & 16438, 17398, 17409, 18427, 18438,
        19877, 20646, 21833, 22831, 24601, 25668 \\ & 25679, 27028, 6161,
        12029, 14269, 15489, 16439, 17399, 17410, 18428, 18439 \\ & 19878, 20647,
        21834, 22832, 24602, 25669, 26381, 27029, 6162,
        13106, 14270 \\ & 15490, 16465, 17400, 17411\\
        
       A2597 & 20811,
        20817,
        20806,
        20805,
        20629,
        20628,
        20627,
        20626,
        19598,
        19597,
        19596 \\ &
        7329,
        6934\\
       
       Phoenix & 19581,
20635,
20634,
16545,
16135,
20631,
20797,
20630,
19583,
20636,
13401 \\ &
19582\\
       
       Perseus & 6146, 6145, 6139, 4953, 4952, 4951, 4949, 4948, 4947, 4946, 4289 \\ & 3209, 11713, 11714, 11715, 11716 \\
       
       MS 0735.6+7421 &  10470,
        10469,
        10822,
        10918,
        10468,
        4197,
        10922,
        10471,
        16275\\
       
       Cygnus A & 17512, 17509, 18871, 5831, 17508, 17514, 17513, 19996, 6225, 17133, \\ & 20079, 17134, 6229, 6250, 20077, 17507, 5830, 6228, 17136, 19989, \\ & 17510, 17511, 18688, 6252, 17135, 6226\\
     \hline
    \end{tabular}
    \caption{The Chandra ObsIDs used for each object.} 
    \label{tab:obsids}
\end{table*}

\section{Additional figures}
\label{app:figures}

\begin{figure*}
    \centering
    \includegraphics[width=0.9\linewidth]{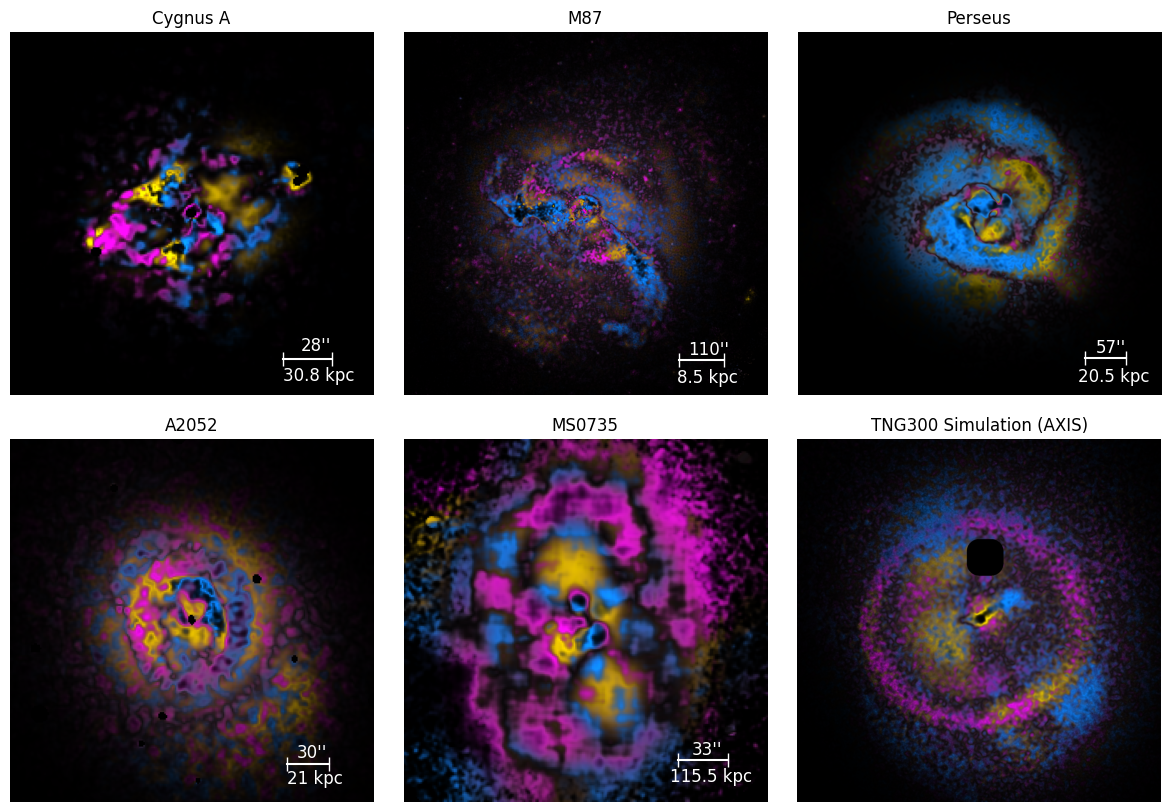}
    \caption{Three-color images of five of the highest significance halos observed with Chandra and the mock AXIS observation of the TNG300 halo. Each perturbation map is assigned a color, such that magenta represents shocked regions, blue represents isobaric regions, and yellow represents bubbles.}
    \label{fig:3color}
\end{figure*}

Fig.~\ref{fig:3color} shows the alternative single image display of the X-arithmetic maps, as mentioned in Section~\ref{sec:methods}, for five of the highest signal-to-noise objects and a TNG300 halo. To produce these images, we consider each pixel of the image. We identify which feature the pixel likely belongs to by requiring that if its value is below a certain threshold in one map (where that feature has been removed), then it also has a value above a certain threshold in the other two maps (where we expect the feature to still be present). The choice of threshold depends on the individual object. If this condition is met, the pixel is colored to match its feature: magenta for shocked regions, blue for isobaric regions, and yellow for bubbles. When this condition is not met, the pixels are black. Only the default version of each object's maps was used to create these images, so some `detections' are spurious, i.e. not present in every version of the maps and therefore not identified in the text.

\begin{figure*}
    \centering
    \includegraphics[width=\textwidth]{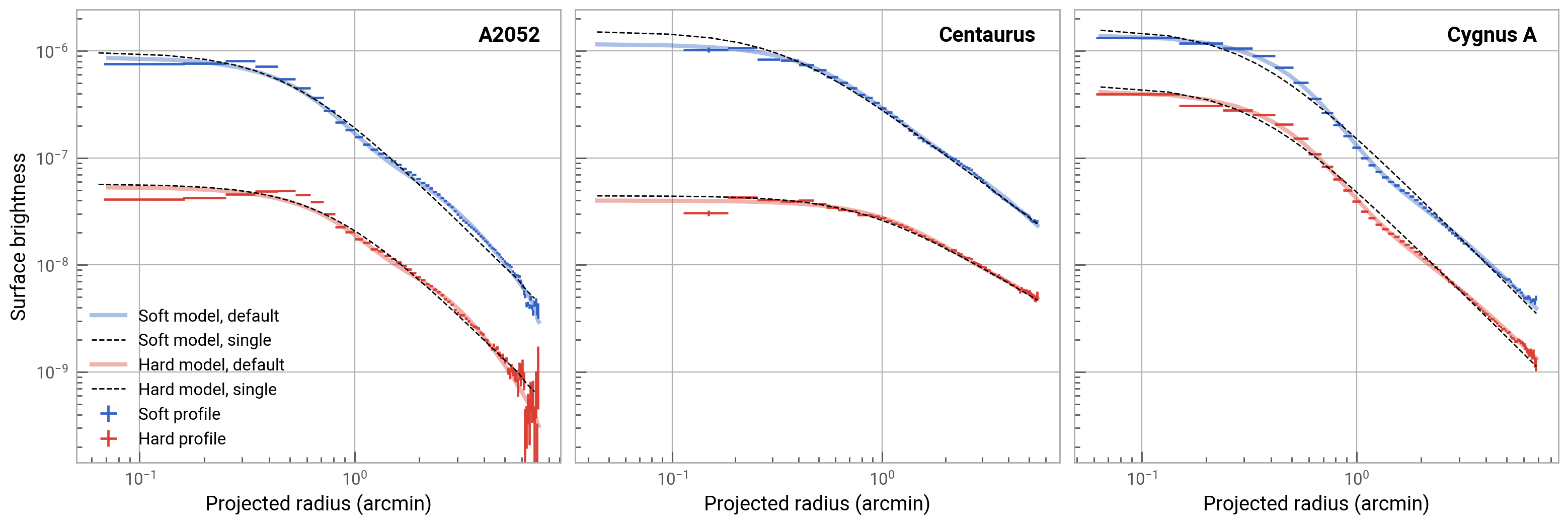}
    \caption{Surface brightness profiles (in arbitrary units) in the soft (blue) and hard (red) bands for (left to right) A2052, Centaurus, and Cygnus A. The best-fit default model type listed in Table~\ref{tab:sample} is shown in light blue (red) for the soft (hard) band. The black dashed lines show the best-fit single spherical or elliptical profiles used in projection calculations as described in Section~\ref{sec:proj}.
    }
    \label{fig:SBprof}
\end{figure*}

A projection correction factor is necessary to account for the different models of the soft and hard band images, as discussed in Section~\ref{sec:proj}. Although a more complicated model was often chosen to produce a residual image, including double $\beta$-models or patching, for simplicity, the best-fit single model to each band was used for the calculation of the projection correction factor. Surface brightness profiles from images in different bands, shown in Fig.~\ref{fig:SBprof}, illustrate the similarity of the single and default models, used for the projection correction and residual image production, respectively.

\clearpage
\bibliography{main}{}
\bibliographystyle{aasjournal}

\end{document}